\documentclass[11pt,journal,final,onecolumn]{IEEEtran}

\usepackage{cite}
\usepackage{epsfig}
\usepackage[english]{babel}
\usepackage{amsmath}
\usepackage{amsthm}
\usepackage{amsfonts}
\usepackage{amssymb}
\usepackage{mdwlist}
\usepackage{graphicx,enumerate,algorithm,algorithmic,multirow,multicol}

\newtheorem{theorem}{Theorem}
\newtheorem{lemma}{Lemma}

\newtheorem{proposition}{Proposition}

\newtheorem{definition}{Definition}

\newcommand{\va}{{{\alpha}}}

\newcommand{\I}{\mathbf{I}}
\newcommand{\vx}{{x}}

\newcommand{\vy}{{y}}

\newcommand{\R}{\mathbb{R}}

\newcommand{\T}{\mathbf{T}}

\newcommand{\be}{\begin{eqnarray}}
\newcommand{\ee}{\end{eqnarray}}

\newcommand{\M}{\mathcal{M}}
\renewcommand{\P}{\mathsf{P}}

\newcommand{\C}{\mathcal{C}}

\newcommand{\card}{\mathrm{Card}~}

\newcommand{\cov}{\mathrm{Cov}}

\newcommand{\norm}[1]{\left\|#1\right\|}
\newcommand{\pds}[2]{\left\langle#1,#2\right\rangle}
\newcommand{\abs}[1]{\left|#1\right|}

\newcommand{\parenth}[1]{\left({#1}\right)}
\newcommand{\crochets}[1]{\left[{#1}\right]}

\newcommand{\eps}{\varepsilon}
\newcommand{\Hm}{\mathcal{H}}
\newcommand{\Tr}{^\mathrm{\small T}}

\renewcommand{\ge}{\geqslant}
\renewcommand{\leq}{\leqslant}

\newcommand{\dft}{f_1}
\newcommand{\reg}{f_2}
\renewcommand{\I}{\mathrm{I}}
\newcommand{\Ad}{\mathrm{A}}

\newcommand{\Hd}{\mathrm{H}}

\newcommand{\Prj}{\mathcal{P}}

\renewcommand{\bf}[1]{\textbf{#1}}

\DeclareMathOperator{\prox}{prox}
\DeclareMathOperator{\rprox}{rprox}
\DeclareMathOperator{\dom}{dom}

\DeclareMathOperator*{\argmin}{arg\ min}

\graphicspath{{images/}}

\title{A proximal iteration for deconvolving Poisson noisy images using sparse
  representations}

\author{F.-X. Dup\'e$^\text{a}$, M.J. Fadili$^\text{a}$ and J.-L. Starck$^\text{b}$
  \begin{tabular}{cc}
    $^\text{a}$ GREYC UMR CNRS 6072 & $^\text{b}$ DAPNIA/SEDI-SAP CEA-Saclay\\
    14050 Caen France & 91191 Gif-sur-Yvette France
  \end{tabular}
}

\begin{document}
%
\maketitle
\begin{abstract}
  We propose an image deconvolution algorithm when the data is contaminated by Poisson
  noise. The image to restore is assumed to be sparsely represented in a dictionary of
  waveforms such as the wavelet or curvelet transforms. Our key contributions are: First, we
  handle the Poisson noise properly by using the Anscombe variance stabilizing transform
  leading to a {\it non-linear} degradation equation with additive Gaussian noise. Second,
  the deconvolution problem is formulated as the minimization of a convex functional with
  a data-fidelity term reflecting the noise properties, and a non-smooth
  sparsity-promoting penalty over the image representation coefficients
  (e.g. $\ell_1$-norm). An additional term is also included in the functional to ensure positivity 
  of the restored image. Third, a fast iterative forward-backward splitting algorithm is
  proposed to solve the minimization problem. We derive existence and uniqueness
  conditions of the solution, and establish convergence of the iterative
  algorithm. Finally, a GCV-based model selection procedure is proposed to objectively select the
  regularization parameter. Experimental results are carried out to show the striking
  benefits gained from taking into account the Poisson statistics of the noise. These
  results also suggest that using sparse-domain regularization may be tractable in many
  deconvolution applications with Poisson noise such as astronomy and microscopy.
\end{abstract}

\newpage

\section{Introduction}
\label{sec:intro}

Deconvolution is a longstanding problem in many areas of signal and image processing
(e.g. biomedical imaging \cite{Sarder2006,Pawley2005}, astronomy \cite{Starck2006}, remote-sensing,
to quote a few). For example, research in astronomical image deconvolution has recently
seen considerable work, partly triggered by the Hubble space telescope (HST) optical
aberration problem at the beginning of its mission. In biomedical imaging, researchers are
also increasingly relying on deconvolution to improve the quality of images acquired by
confocal microscopes \cite{Pawley2005}. Deconvolution may then prove crucial for exploiting images and
extracting scientific content.

There is an extensive literature on deconvolution problems. One might refer to well-known
dedicated monographs on the subject \cite{AndrewsHunt77,Stark87,Jansson97}. In presence of
Poisson noise, several deconvolution methods have been proposed such as Tikhonov-Miller
inverse filter and Richardson-Lucy (RL) algorithms; see \cite{Sarder2006,Starck2006} for a
comprehensive review. The RL has been used extensively in many applications because it is
adapted to Poisson noise. The RL algorithm, however, amplifies noise after a few
iterations, which can be avoided by introducing regularization. In \cite{Dey2004}, the
authors presented a Total Variation (TV)-regularized RL algorithm. In the astronomical
imaging literature, several authors advocated the use of wavelet-regularized RL algorithm
\cite{Starck94,Starck95,Bijaoui04}. In the context of biological imaging deconvolution, wavelets
have also been used as a regularization scheme when deconvolving biomedical images;
\cite{Monvel2001} presents a version of RL combined with wavelets denoising, and
\cite{Vonesch2007} uses the thresholded Landweber iteration introduced in
\cite{Daubechies2004}. The latter approach implicitly assumes that the contaminating noise
is Gaussian.

Other recent attempts for solving Poisson linear inverse problems is a Bayesian multi-scale framework proposed in \cite{NowakKolaczyk} based on a multi-scale factorization of the Poisson likelihood function associated with a recursive partitioning of the underlying intensity. Regularization of the solution is accomplished by imposing prior probability distributions in a Bayesian paradigm and the maximum a posteriori solution is computed using the expectation-maximization algorithm. However, the multiscale analysis by the above authors is only tractable with the Haar wavelet.
Similarly, the work in \cite{CavalierKoo} on hard threshold estimators in the tomographic data framework has shown that for a particular operator (the Radon operator) an extension of wavelet-vaguelette decomposition (WVD) method \cite{DonohoWVD} for Poisson data is theoretically feasible. But the authors do not provide any computational algorithm for computing the estimate.
Inspired by the WVD method, the authors in \cite{AntoniadisBigot} explored an alternative approach via wavelet-based decompositions combined with thresholding strategies that address adaptivity issues. Specifically, their framework extends the wavelet-Galerkin methods of \cite{Cohen04} to the Poisson setting. In order to ensure the positivity of the estimated intensity, the log-intensity is expanded in a wavelet basis. This method is however limited to standard orthogonal wavelet bases.

In the context of deconvolution with Gaussian white noise, sparsity-promoting
regularization over orthogonal wavelet coefficients has been recently proposed
\cite{Figueiredo2003,Daubechies2004,Combettes2005}. Generalization to frames was proposed
in \cite{Teschke2007,Combettes2007b}. In \cite{Fadili2006a}, the authors presented an
image deconvolution algorithm by iterative thresholding in an overcomplete dictionary of
transforms, and \cite{Starck03} designed a deconvolution method that combines both the
wavelet and curvelet transforms.
However, sparsity-based approaches published so far have mainly focused on Gaussian noise.

In this paper, we propose an image deconvolution algorithm for data blurred and
contaminated by Poisson noise. The Poisson noise is handled properly by using the Anscombe
variance stabilizing transform (VST), leading to a {\it non-linear} degradation equation
with additive Gaussian noise, see \eqref{eq:3}. The deconvolution problem is then
formulated as the minimization of a convex functional combining a non-linear data-fidelity term
reflecting the noise properties, and a non-smooth sparsity-promoting penalty over the
representation coefficients of the image to restore. Such representations include not only the orthogonal wavelet transform 
but also overcomplete representations such as translation-invariant wavelets, curvelets or wavelets and curvelets.
Since Poisson intensity functions are nonnegative by definition, an additional term is also included in the minimized functional
to ensure the positivity of the restored image. Inspired by the work in \cite{Combettes2005}, a fast proximal iterative
algorithm is proposed to solve the minimization problem. Experimental results are carried
out on a set of simulated and real images to compare our approach to some competitors. We
show the striking benefits gained from taking into account the Poisson nature of the noise
and the morphological structures involved in the image through overcomplete sparse
multiscale transforms.

\subsection{Relation to prior work}
\label{sec:relation-prior-work}

A naive solution to this deconvolution problem would be to apply traditional approaches
designed for Gaussian noise. But this would be awkward as (i) the noise tends to Gaussian
only for large mean intensities (central limit theorem), and (ii) the noise variance
depends on the mean anyway. A more adapted way would be to adopt a bayesian framework with
an appropriate anti-log-likelihood score{\textemdash}the negative of the log-likelihood function{\textemdash} to obtain a data fidelity term reflecting the Poisson statistics of the noise. The data fidelity term is derived from the conditional distribution of the observed data given the original image, which is known to be governed by physical considerations concerned with the data-acquisition device and the noise generating process (e.g. Poisson here). Unfortunately, doing so, we would end-up with a functional which does not satisfy a key property: the data fidelity term does not have a Lipschitz-continuous gradient as required in \cite{Combettes2005}, hence preventing us
from using the forward-backward splitting proximal algorithm to solve the optimization
problem. To circumvent this difficulty, we propose to handle the noise statistical
properties by using the Anscombe VST. Some previous authors \cite{ChauxSPIE} have already
suggested to use the Anscombe VST, and then deconvolve with wavelet-domain regularization
as if the stabilized observation were linearly degraded and contaminated by additive
Gaussian noise. But this is not valid as standard results of the Anscombe VST
lead to a non-linear degradation equation because of the square-root, see \eqref{eq:3}.

\subsection{Organization of this paper}
\label{sec:organ-this-paper}

The organization of the paper is as follows: we first formulate our deconvolution problem
under Poisson noise (Section \ref{sec:problem-statement}), and then recall some necessary
material about overcomplete sparse representations (Section
\ref{sec:sparse-image-repr}). The core of the paper lies in Section
\ref{sec:sparse-iter-deconv}, where we state the deconvolution optimization problem,
characterize it and solve it using monotone operator splitting iterations. We also focus
on the choice of the two main parameters of the algorithm and propose some solutions. In
Section \ref{sec:results}, experimental results are reported and discussed. The proofs of
our main results are deferred to the appendix for the sake of presentation.

\subsection{Notation and terminology}
\label{sec:notation}

Let $\Hm$ a real Hilbert space, here a finite dimensional real vector space. We
denote by $\norm{.}$ the norm associated with the inner product $\pds{.}{.}$ in $\Hm$, and
$\I$ is the identity operator on $\Hm$. $\vx$ and $\va$ are respectively reordered vectors
of image samples and transform coefficients. 

A real-valued function $f$ is coercive, if $\lim_{\norm{\vx} \to +\infty}f\parenth{\vx}=+\infty$. 
The domain of $f$ is defined by $\dom f = \{ x\in\Hm\ :\ f(x) < +\infty \}$ and $f$ is proper if $\dom f \neq
\emptyset$. We say that a real-valued function $f$ is lower
semi-continuous (lsc) if $\liminf_{x \to x_0} f(x) \geq f(x_0)$. 
Lower semi-continuity is weaker than continuity, and plays an important role for existence 
of solutions in minimization problems \cite[Page 17]{LemarechalHiriart96}. $\Gamma_0(\Hm)$ is the class of all 
proper lsc convex functions from $\Hm$ to $(-\infty,+\infty]$. The subdifferential of a function $f \in \Gamma_0(\Hm)$ at $x \in \Hm$
is the set $\partial f(x) = \left\{u \in \Hm | \forall y \in \Hm, f(y) \geq f(x) + \pds{u}{y-x}\right\}$. An element $u$ of $\partial f$ is called a subgradient. A comprehensive account of subdifferentials can be found in \cite{LemarechalHiriart96}.

An operator $\mathrm{A}$ acting on $\Hm$ is $\kappa$-Lipschitz continuous if
$\forall x,y \in \Hm, \norm{\mathrm{A}(x) - \mathrm{A}(y)} \leq \kappa \norm{x - y}$ where
$\kappa$ is the Lipschitz constant. The spectral operator norm of $\mathrm{A}$ is given by $\norm{\mathrm{A}}_2= \max_{\vx \neq 0} \frac{\norm{\mathrm{A}\vx}}{\norm{\vx}}$. 

We denote by $\imath_{\C}$ the indicator of the convex set $\C$: $ \imath_{\C} (x) =
  \begin{cases}
    0, & \text{if } x \in \C ~ ,\\
    +\infty, & \text{otherwise}.
  \end{cases}$.
We denote by $\rightharpoonup$ the convergence.

\section{Problem statement}
\label{sec:problem-statement}

Consider the image formation model where an input image of $n$ pixels $\vx$ is blurred by
a point spread function (PSF) $h$ and contaminated by Poisson noise. The observed image is
then a discrete collection of counts $\vy=(y_i)_{1 \leq i \leq n}$ which are bounded,
i.e. $\vy \in \ell_{\infty}$.  Each count $y_i$ is a realization of an independent Poisson
random variable with a mean $(h \circledast x)_i$, where $\circledast$ is the circular
convolution operator. Formally, this writes $y_i \sim \Prj\parenth{(h \circledast x)_i}$. 
The deconvolution problem at hand is to restore $\vx$ from the observed count image $\vy$.

A natural way to attack this problem would be to adopt a maximum a posteriori (MAP)
bayesian framework with an appropriate likelihood function{\textemdash}the distribution of the observed data $\vy$ given 
an original $\vx${\textemdash}reflecting the Poisson statistics of the noise. But, as stated above, this would prevent
us from using the forward-backward splitting proximal algorithm to solve the MAP
optimization problem, since the gradient of the data fidelity term is not
Lipschitz-continuous. Indeed, forward-backward iteration is essentially a generalization of the classical gradient projection method \cite{Ciarlet85} for constrained convex optimization and monotone variational inequalities, and inherit restrictions similar to those methods. For such methods, Lipschitz continuity of the gradient is classical \cite[Theorem 8.6-2]{Ciarlet85}. The latter property is then crucial for the iterates in \eqref{eq:5} to be determined uniquely, and for the forward-backward splitting algorithm to converge; see Theorem~\ref{th:2} and also \cite{Combettes04}. For this reason, we propose to handle the noise statistical properties by using the Anscombe VST \cite{Anscombe48} defined as
\begin{equation}
\label{eq:3}
z_i = 2\sqrt{(h\circledast x)_i+\tfrac{3}{8}} + \eps,\quad \eps \sim \mathcal{N}(0,1),
\end{equation}
where $\eps$ is an additive white Gaussian noise of unit variance\footnote{Rigorously 
speaking, the equation is to be understood in an asymptotic sense.}. In words, $z$ is
{\em non-linearly} related to $x$. In Section~\ref{sec:sparse-iter-deconv}, we provide an
elegant optimization problem and a fixed point algorithm taking into account such a
non-linearity.

\section{Sparse image representation}
\label{sec:sparse-image-repr}

Let $x \in \Hm$ be an $\sqrt{n}\times\sqrt{n}$ image. $x$ can be written as the
superposition of elementary atoms $\varphi_\gamma$ parameterized by $\gamma \in
\mathcal{I}$ according to the following linear generative model :
\begin{equation}
  \label{eq:4}
  x = \sum_{\gamma \in \mathcal{I}} \va_\gamma \varphi_\gamma = \Phi \va,
  \quad \abs{\mathcal{I}} = L \ge n ~ .
\end{equation}
We denote by $\Phi$ the dictionary i.e. the $n\times L$ matrix whose columns are the
generating waveforms $\parenth{\varphi_\gamma}_{\gamma \in \mathcal{I}}$ all normalized to
a unit $\ell_2$-norm. The forward (analysis) transform is then defined by a
non-necessarily square matrix $\T = \Phi\Tr \in \mathbb{R}^{L\times n}$ with $L\ge
n$. When $L > n$ the dictionary is said to be redundant or overcomplete. In the case of
the simple orthogonal basis, the inverse (synthesis) transform is trivially $\Phi =
\T\Tr$. Whereas assuming that $\Phi$ is a tight frame implies that the frame operator
satisfies $\Phi\Phi\Tr = c\I$, where $c > 0$ is the tight frame constant. For tight
frames, the pseudo-inverse reconstruction (synthesis) operator reduces to $c^{-1}\Phi$. In
the sequel, the dictionary $\Phi$ will correspond either to an orthobasis or to a tight frame of
$\Hm$.

Owing to recent advances in modern harmonic analysis, many redundant systems, like the
undecimated wavelet transform, curvelet, contourlet etc, were shown to be very effective in
sparsely representing images. By sparsity, we mean that we are seeking for a good
representation of $x$ with only few significant coefficients. 

In the rest of the paper, the dictionary $\Phi$ is built by taking union of one or several
transforms, each corresponding to an orthogonal basis or a tight
frame. Choosing an appropriate dictionary is a key step towards a good sparse
representation, hence restoration. A core idea here is the concept of morphological
diversity. When the transforms are amalgamated in the dictionary, they have to be chosen
such that each leads to sparse representation over the parts of the image it is serving 
while being inefficient in representing the other image content. As popular examples, one may think of
wavelets for smooth images with isotropic singularities \cite[Section 9.3]{Mallat98}, curvelets
for representing piecewise smooth $C^2$ images away from $C^2$ contours
\cite{CandesDonohoCurvelets,CandesFDCT05}, wave atoms or local DCT to represent
locally oscillating textures \cite{Demanet06,Mallat98}. 

\section{Sparse Iterative Deconvolution}
\label{sec:sparse-iter-deconv}

\subsection{Optimization problem}
\label{sec:sparse-optim}

In this Section, we derive that the class of minimization problems we are interested in, see \eqref{eq:9}, can be stated in the general form :
\begin{equation}
  \label{eq:6}
  \min_{\va \in \R^L} \dft(\va) + \reg(\va),
\end{equation}
where $\dft \in \Gamma_0(\R^L)$, $\reg\in\Gamma_0(\R^L)$
and $\dft$ is differentiable with a $\kappa$-Lipschitz
gradient. We denote by $\M$ the set of solutions of \eqref{eq:6}.

\noindent
From \eqref{eq:3}, we immediately deduce the data fidelity term
\begin{gather}
\label{eq:7}
  F \circ \Hd \circ \Phi~(\va), ~ \text{with} \\
  F : \eta \in \mathbb{R}^n \mapsto \sum_{i=1}^n f(\eta_i),\quad f(\eta_i) = \frac{1}{2}
  \left( z_i - 2\sqrt{\eta_i+\tfrac{3}{8}} \right)^2 , \nonumber 
\end{gather}
where $\Hd$ denotes the (circular) convolution operator. From a statistical
perspective, \eqref{eq:7} corresponds to the anti-log-likelihood score. Note that for bias correction reasons \cite{Jansen06}, the value 1/8 may be used instead of 3/8 in \eqref{eq:7}. However, there are implications of this alternate choice on the Lipschitz constant in \eqref{eq:22}, and consequently it can be seen from Theorem \ref{th:2} that this will have an unfavorable impact on the convergence speed of the deconvolution algorithm.

Adopting a bayesian framework and using a standard MAP rule, our
goal is to minimize the following functional with respect to the representation
coefficients $\va$ :
\begin{gather}
  \label{eq:9}
  (\P_{\lambda,\psi}): \min_{\va} J(\va) ~ ,\\
  J : \va \mapsto \underbrace{F \circ \Hd \circ \Phi\ (\va)}_{\dft(\va)} + 
  \underbrace{\imath_{\C} \circ \Phi\ (\va) + \lambda \sum_{i=1}^L \psi(\va_i)}_{\reg(\va)}, \nonumber 
\end{gather}
where we implicitly assumed that $(\va_i)_{1 \leq i \leq L}$ are independent and
identically distributed with a Gibbsian density $\propto e^{-\lambda\psi(\va_i)}$. 
The penalty function $\psi$ is chosen to enforce sparsity, $\lambda > 0$ is a regularization
parameter and $\imath_{\C}$ is the indicator function of the convex set $\C$. In our case,
$\C$ is the positive orthant. The role of the term $\imath_{\C} \circ \Phi$ is to impose the positivity 
constraint on the restored image because we are fitting Poisson intensities, which are positive by nature. 
We also define the set $\C'=\{\va | \Phi\va \in \C\}$, that is $\imath_{\C'}=\imath_{\C} \circ \Phi$.

From \eqref{eq:9}, we have the following,
\begin{proposition}
  \label{prop:1}
  {~} \\\vspace*{-0.5cm}
  \begin{enumerate}[(i)]
  \item $\dft$ is convex function. It is strictly convex if $\Phi$ is an orthobasis and
    $\mathrm{ker}\parenth{\Hd}=\emptyset$ (i.e. the spectrum of the PSF does not vanish
    within the Nyquist band).
  \item The gradient of $\dft$ is
    \begin{equation}
      \label{eq:19}
      \nabla \dft (\va) = \Phi\Tr \circ \Hd\Tr \circ \nabla F \circ \Hd \circ \Phi\ (\va) ~ ,
    \end{equation}
    with
    \begin{equation}
      \label{eq:20}
      \nabla F (\eta) = \parenth{\frac{-z_i}{\sqrt{\eta_i + 3/8}} + 2}_{1\leq i\leq n} ~ .
    \end{equation}
  \item $\dft$ is continuously differentiable with a $\kappa$-Lipschitz gradient where
    \begin{equation}
      \label{eq:22}
      \kappa \leq \parenth{\tfrac{2}{3}}^{3/2} 4 c \norm{\Hd}_2^2 \norm{z}_{\infty} < +\infty.
    \end{equation}
  \item $(\P_{\lambda,\psi})$ is a particular case of problem \eqref{eq:6}.
  \end{enumerate}   
\end{proposition}
A proof can be found in the appendix.

\subsection{Characterization of the solution}
\label{sec:char-solut}

Since $J$ is coercive and convex, the following holds :
\begin{proposition}
  \label{prop:2}
  {~} \\\vspace{-0.5cm}
  \begin{enumerate*}
  \item Existence: $(\P_{\lambda,\psi})$ has at least one solution, i.e. $\M \neq
    \emptyset$.
  \item Uniqueness: $(\P_{\lambda,\psi})$ has a unique solution if $\Phi$ is an orthobasis
    and $\mathrm{ker}\parenth{\Hd}=\emptyset$, or if $\psi$ is strictly convex.
  \end{enumerate*}
\end{proposition}

\subsection{Proximal iteration}
\label{sec:proximal-iteration}
We first define the notion of a proximity operator, which was introduced in
\cite{Moreau1962} as a generalization of the notion of a convex projection operator.
\begin{definition}[Moreau\cite{Moreau1962}]
  \label{def:1} 
  Let $\varphi \in \Gamma_{0}(\Hm)$. Then, for every $x\in\Hm$, the
  function $y \mapsto \varphi(y) + \norm{x-y}^{2}/2$ achieves its infimum at a unique
  point denoted by $\prox_{\varphi}x$. The operator $\prox_{\varphi} : \Hm \to \Hm$ thus
  defined is the \textit{proximity operator} of $\varphi$.
  Moreover, $\forall x,p \in \Hm$
  \begin{align}
    \label{eq:prox}
    p = \prox_{\varphi} x \iff x-p \in \partial\varphi(p) \iff \pds{y-p}{x-p} + \varphi(p) \leq \varphi(y) ~ \forall y\in\Hm.
  \end{align}
  \eqref{eq:prox} means that $\prox_{\varphi}$ is the resolvent of the subdifferential of $\varphi$ \cite{Eckstein92}. Recall that the resolvent of the subdifferential $\partial\varphi$ is the single-valued operator $J_{\partial\varphi}: \Hm \to \Hm$ such that $\forall x \in \Hm, x - J_{\partial\varphi}(x) \in \partial\varphi(J_{\partial\varphi}) \iff J_{\partial\varphi} = (\I - \partial\varphi)^{-1}$.
   
\end{definition}
\noindent
It will also be convenient to introduce the reflection operator $\rprox_{\varphi} = 2\prox_{\varphi} - \I$.\\

For notational simplicity, we denote by $\Psi$ the function $\va \mapsto \sum_i
\psi(\va_i)$. Our goal now is to express the proximity operator associated to $\reg$,
which will be needed in the iterative deconvolution algorithm. The difficulty stems from
the definition of $\reg$ which combines both the 'positivity' constraint and the
regularization. Unfortunately, we can show that even with a separable penalty $\Psi(\va)$,
the operator $\prox_{f_2}=\prox_{\imath_{\C} \circ \Phi + \lambda \Psi}$ has no explicit
form in general, except the case where $\Phi = \I$. We then propose to replace explicit
evaluation of $\prox_{f_2}$ by a sequence of calculations that activate separately
$\prox_{\imath_{\C} \circ \Phi}$ and $\prox_{\lambda \Psi}$. We will show that the last
two proximity operators have closed-form expressions. Such a strategy is known as a
splitting method of maximal monotone operators \cite{LionsMercier79,Eckstein92}. As both
$\imath_{\C'}$ and $\Psi$ belong to $\Gamma_0\parenth{\Hm}$ and are non-differentiable, our splitting
method is based on the Douglas-Rachford algorithm \cite{LionsMercier79,Eckstein92,Combettes04}. The
following lemma summarizes our scheme.

\begin{lemma}
  \label{lemma:2}
  Let $\Phi$ an orthobasis or a tight frame with constant $c$. Recall that $\C'=\{\va | \Phi\va \in \C\}$.
  \begin{enumerate}
  \item If $\va \in \C'$ then $\prox_{f_2}(\va) = \prox_{\lambda\Psi}(\va)$.
  \item Otherwise, let $(\nu_t)_t$ be a sequence in $(0,1)$ such that $\sum_t \nu_t(1-\nu_t)=+\infty$. Take $\gamma^{0} \in \Hm$, and
    define the sequence of iterates :
    \begin{eqnarray}
      \label{eq:proxtframe1}
      \gamma^{t+1} = \gamma^t + \nu_t\parenth{\rprox_{\lambda\Psi + \tfrac{1}{2}\norm{. - \va}^2}\circ\rprox_{\imath_{\C'}} - \I}(\gamma^t) ,
\end{eqnarray}
where $\prox_{\lambda\Psi + \tfrac{1}{2}\norm{. - \va}^2}
(\gamma^t)= \parenth{\prox_{\tfrac{\lambda}{2}\psi} \parenth{(\va_i + \gamma^t_i)/2}}_{1\leq i \leq
  L}$, $\Prj_{\C'} = \prox_{\imath_{\C'}} = c^{-1} \Phi\Tr\circ\Prj_\C\circ\Phi
+\parenth{\I - c^{-1} \Phi\Tr\circ\Phi}$ and $\Prj_\C$ is the projector onto the positive
orthant $(\Prj_\C \eta)_i = \max(\eta_i, 0)$. Then,
   \begin{equation}
    \label{eq:proxtframe2}
    \gamma^t \rightharpoonup \gamma ~ \text{and} ~ \prox_{\reg}(\va) = \Prj_{\C'}(\gamma).
   \end{equation}
  \end{enumerate}
\end{lemma}
The proof is detailed in the appendix. Note that when $\Phi$ is an orthobasis,
$\Prj_{\C'}=\Phi\Tr\circ\Prj_\C\circ\Phi$.

\noindent
To implement the above iteration, we need to express $\prox_{\lambda\psi}$, which is given
by the following result for a wide class of penalties $\psi$ :
\begin{lemma}
\label{th:3}
Suppose that $\psi$ satisfies, (i) $\psi$ is convex even-symmetric , non-negative and
non-decreasing on $[0,+\infty)$, and $\psi(0)=0$. (ii) $\psi$ is twice differentiable on
$\mathbb{R}\setminus \{0\}$. (iii) $\psi$ is continuous on $\mathbb{R}$, it is not
necessarily smooth at zero and admits a positive right derivative at zero $\psi^{'}_+(0) =
\lim_{h\to 0^+} \frac{\psi(h)}{h} > 0$. Then, the proximity operator of $\delta\Psi(\gamma)$,
$\prox_{\delta\Psi}(\gamma)$ has exactly one continuous solution
decoupled in each coordinate $\gamma_i$ :
\begin{equation}
  \label{eq:10}
  \prox_{\delta\psi}(\gamma_i) =
  \begin{cases}
    0 & \text{if } \abs{\gamma_i} \leq \delta\psi^{'}_+(0)\\
    \gamma_i-\delta\psi^{'}(\bar{\va}_i) & \text{if } \abs{\gamma_i} > \delta\psi^{'}_+(0)
  \end{cases}
\end{equation}
\end{lemma}
A proof of this lemma can be found in \cite{Fadili2006}. A similar result also has
recently appeared in \cite{Combettes2007}. Among the most popular penalty functions $\psi$
satisfying the above requirements, we have $\psi(\va_i) = \abs{\va_i}$, in which case the
associated proximity operator is the popular soft-thresholding. 

\noindent
We are now ready to state our main proximal iterative algorithm to solve the minimization
problem $(\P_{\lambda,\psi})$:
\begin{theorem}
  \label{th:2}
  For $t \geq 0$, let $(\mu_{t} )_{t}$ be a sequence in $(0, +\infty)$ such that $0 <
  \inf_{t} \mu_{t} \leq \sup_{t} \mu_{t}< \parenth{\frac{3}{2}}^{3/2}/\parenth{2 c
    \norm{\Hd}_2^2 \norm{z}_{\infty}}$, let $(\beta_t)_t$ be a sequence in $(0,1]$ such that
  $\inf_t \beta_t > 0$, and let $(a_t)_t$ and $(b_t)_t$ be sequences in $\Hm$ such
  that $\sum_t \norm{a_t} < +\infty$ and $\sum_t \norm{b_t} < +\infty$. Fix
  $\va^{0}\in\R^L$, for every $t \ge 0$, set
  \begin{equation}
    \label{eq:5}
    \va^{t+1} = \va^t + \beta_t ( \prox_{\mu_{t}f_{2}}\parenth{\va^{t} -
      \mu_{t}\parenth{ \nabla f_{1}(\va^{t}) + b_t }} + a_t - \va^t)
  \end{equation}
  where $\nabla f_{1}$ and $\prox_{\mu_{t}f_{2}}$ are given by Proposition \ref{prop:1}(ii) and
  Lemma \ref{lemma:2}.
  \noindent 
  Then $(\va_t)_{t \geq 0}$ converges to a solution of $(\P_{\lambda,\psi})$.
\end{theorem}
This is the most general convergence result known on the forward-backward iteration. 
The name of the iteration is inspired by well-established techniques from numerical linear algebra.
The words "forward" and "backward" refer respectively to the standard notions of a forward difference (here explicit gradient descent) 
step and of a backward difference (here implicit proximity) step in numerical analysis. The
sequences $a_t$ and $b_t$ play a prominent role as they formally establish the robustness of the algorithm to
numerical errors when computing the gradient $\nabla f_{1}$ and the proximity operator
$\prox_{\reg}$. The latter remark will allow us to accelerate the algorithm 
by running the sub-iteration \eqref{eq:proxtframe1} only a few iterations
(see implementation details in \ref{sec:comp-compl-impl}).

For illustration, let's take $\Psi$ as the $\ell_1$ norm, in which case $\prox_{\lambda\Psi}$ is the component-wise soft-thresholding with threshold $\lambda$, $a_t = b_t \equiv 0$, $\beta_t \equiv 1$ and $\mu_t \equiv \mu$ in \eqref{eq:5}, and $\nu_t \equiv 1/2$ in \eqref{eq:proxtframe1}. Thus, bringing all the pieces together, the deconvolution algorithm dictated by iterations \eqref{eq:5} and \eqref{eq:proxtframe1} is summarized in Algorithm~\ref{algo:1}.

\begin{algorithm}
\noindent{\bf{Task:}} Image deconvolution with Poisson noise, solve \eqref{eq:9}. \\
\noindent{\bf{Parameters:}} The observed image counts $y$, the dictionary $\Phi$, number of iterations $N_{\mathrm{FB}}$ in \eqref{eq:5} and $N_{\mathrm{DR}}$ in sub-iteration \eqref{eq:proxtframe1}, relaxation parameter $\mu$, regularization parameter $\lambda$.\\
\noindent{\bf{Initialization:}}
\begin{itemize}
\item Apply VST $z=2\sqrt{y+3/8}$.
\item Initial solution $\va^{0} = 0$.
\end{itemize}
\noindent{\bf{Main iteration:}} \\
\noindent{\bf{For}} $t=0$ {\bf{to}} $N_{\mathrm{FB}}-1$,
\begin{itemize}
\item Compute blurred estimate $\eta^{t} = \Hd\Phi\alpha^{t}$.
\item Compute 'residuals' $r^{t} = \Big(\frac{-z_i}{\sqrt{\eta^{t}_i + 3/8}} + 2\Big)_{1\leq i\leq n}$.
\item Move along the descent direction $\xi^{t}=\alpha^{t} + \mu\Phi\Tr\Hd\Tr r^{t}$.
\item Initialize $\gamma^{0} = \xi^{t}$, and start sub-iteration.
\item {\bf For} $m=0$ {\bf{to}} $N_{\mathrm{DR}}$-1,
\begin{itemize}
\item Project $\gamma^{m}$ orthogonally to $\C'$: $\zeta^{m} = c^{-1}\Phi\Tr\parenth{\min(\Phi \gamma^{m},0)}$.
\item Update $\gamma^{m+1}$ by soft-thresholding $\gamma^{m+1} = \mathrm{ST_{\lambda/2}}\parenth{(\xi^{t}+\gamma^{m})/2 - \zeta^{m}} + \zeta^m$.
\end{itemize}
\item Update $\alpha^{t+1} = \gamma^{N_{\mathrm{DR}}} - c^{-1}\Phi\Tr\parenth{\min(\Phi \gamma^{N_{\mathrm{DR}}},0)}$.
\end{itemize}
\noindent{\bf{End main iteration}} \\
\noindent{\bf{Output:}} Deconvolved image $x^{\star}=\Phi\alpha^{N_{\mathrm{FB}}}$.
\caption{}
\label{algo:1}
\end{algorithm}

\subsection{Choice of $\mu$}
\label{sec:choice-mu}

The relaxation (or descent) parameter $\mu$ has an important impact on the convergence speed of the
algorithm. The upper-bound provided by Theorem \ref{th:2}, derived from the
Lipschitz constant \eqref{eq:22} is only a sufficient condition for \eqref{eq:5} to converge, 
and may be pessimistic in some applications. To circumvent this drawback, Tseng proposed in \cite{Tseng2000} 
an extension of the forward-backward algorithm with an iteration to adaptively estimate a "good" value of $\mu$. 
The main result provided hereafter is an adaptation to our context to the one of Tseng
\cite{Tseng2000}. We state it in full for the sake of completeness and the reader
convenience.

\begin{theorem}
  Let $\C'$ as defined above (\ref{sec:sparse-optim}). Choose any $\va_0 \in \C'$.  Let $(\mu_t)_{t \in \mathbb{N}}$
  be a sequence such that $\forall t > 0, \mu_t \in (0,\infty)$. Let $\dft$
  as defined in \eqref{eq:9}. Then the sequence $(\va_t)_{t\in\mathbb{N}}$ of iterates
    \begin{equation}
    \label{eq:tseng}
    \begin{split}
      \va_{t+\tfrac{1}{2}} &= \prox_{\lambda\Psi} \parenth{ \va_t - \mu_t \nabla \dft (\va_t)} ~ ,\\
      \va_{t+1} &= \Prj_{\C'} \parenth{\va_{t+\tfrac{1}{2}} - \mu_t
      \parenth{\nabla \dft(\va_{t+\tfrac{1}{2}}) - \nabla \dft (\va_t)}}
    \end{split}
    \end{equation}
    converges to a minimum of $J$.
\end{theorem}
As $\nabla\dft$ is Lipschitz-continuous, the update of the relaxation sequence $\mu_t$ is rather easy. Indeed, using an
Armijo-Goldstein-type stepsize approach, we can compute and update $\mu_t$ at each
iteration by taking $\mu_t$ to be the largest $\mu \in
\{\sigma,\tau\sigma,\tau^2\sigma,\ldots\}$ satisfying
\begin{equation}
  \label{eq:11}
  \mu\norm{\nabla \dft(\va_{t+\tfrac{1}{2}}) - \nabla \dft(\va_t)} \leq \theta
  \norm{\va_{t+\tfrac{1}{2}} - \va_t} ~ ,
\end{equation}
where $\tau\in(0,1)$, $\theta\in(0,1)$ and $\sigma > 0$ are constants. $\tau=1/2$ is a typical choice.


It is worth noting that for tight frames, this algorithm will somewhat simplify the
computation of $\prox_{f_2}$, removing the need of the Douglas-Rachford sub-iteration
\eqref{eq:proxtframe1}. But, whatever the transform, this will come at the price of
keeping track of the gradient of $\dft$ at the points $\va_{t+\tfrac{1}{2}}$ and $\va_{t}$,
and the need to check \eqref{eq:11} several times.

\subsection{Choice of $\lambda$}
\label{sec:choice-lambda}

As usual in regularized inverse problems, the choice of $\lambda$ is crucial as it
represents the desired balance between sparsity (regularization) and deconvolution (data
fidelity). For a given application and corpus of images (e.g. confocal microscopy), a
naive brute-force approach would consist in testing several values of $\lambda$ and taking
the best one by visual assessment of the deconvolution quality. However, this is
cumbersome in the general case.

We propose to objectively select the regularizing parameter $\lambda$ based on an adaptive model selection criterion
such as the generalized cross validation (GCV) \cite{Golub1979}. Other criteria are possible as well including the AIC \cite{Akaike73} or the BIC \cite{Schwarz78}. GCV attempts to provide a data-driven estimate of $\lambda$ by minimizing :
\begin{equation}
  \label{eq:gcv1}
  \mathrm{GCV}(\lambda) = \frac{\norm{z - 2\sqrt{\Hd\Phi\va^{\star} + \frac{3}{8}}}^2}{(n - df)^2},
\end{equation}
where $\va^{\star}(z)$ denotes the solution arrived at by iteration \eqref{eq:5} (or \eqref{eq:tseng}), 
and $df$ is the effective number of degrees of freedom.

Deriving the closed-form expression of $df$ is very challenging in our case as it faces two
main difficulties, (i) the observation model \eqref{eq:3} is non-linear, and (ii) the
solution $\alpha^{\star}(z)$ is not known in closed form but given by the iterative forward-backward
algorithm. 

Degrees of freedom is a familiar phrase in statistics. In (overdetermined) linear regression $df$ is the number of estimated predictors. More generally, degrees of freedom is often used to quantify the model complexity of a statistical modeling procedure. However, generally speaking, there is no exact correspondence between the degrees of freedom $df$ and the number of parameters in the model. In penalized solutions of inverse problems where the estimator is linear in the observation, e.g. Tikhonov regularization or ridge regression in statistics, $df$ is simply the trace of the so-called influence or the hat matrix. But in general, it is difficult to derive the analytical expression of $df$ for general nonlinear modeling procedures such as ours. This remains a challenging and active area of research. 

Stein's unbiased risk estimation (SURE) theory \cite{Stein81} gives a rigorous definition of the degrees of freedom for any fitting procedure. Following our notation, given the solution $\alpha^{\star}$ provided by our deconvolution algorithm, let $z^{\star}(z)=2\sqrt{\Hd\Phi\va^{\star}(z)+3/8}$ represent the model fit from the observation $z$. As $Z|\alpha \sim \mathcal{N}\parenth{2\sqrt{\Hd\Phi\alpha + 3/8},1}$, it follows from \cite{Efron86} that the degrees of freedom of our procedure is 
\[
df(\lambda) = \sum_{i=1}^n\cov(z_i^{\star}(z),z_i) ~,
\]
a quantity also called the optimism of the estimator $z^{\star}(z)$. If the estimation algorithm is such that $z^{\star}(z)$ is almost-differentiable \cite{Stein81} with respect to $z$, so that its divergence is well-defined in the weak sense (as is the case if $z^{\star}(z)$ were uniformly Lipschitz-continuous), Stein's Lemma yields the so-called divergence formula
\be
\label{eq:dfdiv}
df(\lambda) = \sum_{i=1}^n\cov(z_i^{\star}(z),z_i) = \mathrm{E}_Z\left[\mathrm{div}\parenth{z^{\star}(z)}\right] = \mathrm{E}_Z\left[\sum_{i=1}^n \frac{\partial z_i^{\star}(z)}{\partial z_i}\right] ~.
\ee
where the expectation $\mathrm{E}_Z$ is taken under the distribution of $Z$. The $df$ is then the sum of the sensitivity of each fitted value with respect to the corresponding observed value. For example, the last expression of this formula has been used in \cite{Jansen97} for orthogonal wavelet denoising. However, it is notoriously difficult to derive the closed-form analytical expression of $df$ from the above formula for general nonlinear modeling procedures. To overcome the analytical difficulty, the bootstrap \cite{Efron04} can be used to obtain an (asymptotically) unbiased estimator of $df$. Ye \cite{Ye98} and Shen and Ye \cite{Shen02} proposed using a data perturbation technique to numerically compute an (approximately) unbiased estimate for $df$ when the analytical form of $z^{\star}(z)$ is unavailable. From \eqref{eq:dfdiv}, the estimator of $df$ takes the form
\begin{eqnarray}
\widehat{df}(\lambda) &=& \mathrm{E}_{V_0}\crochets{\pds{v_0}{\frac{z^{\star}(z+\tau v_0) - z^{\star}(z)}{\tau}}} ,~ V_0 \sim \mathcal{N}(0,\mathrm{I}) ~ ,\nonumber \\
		      &=& \frac{1}{\tau^2}\int \pds{v}{z^{\star}(z+v)} \phi(v;\tau^2\mathrm{I}) dv, ~ V \sim \mathcal{N}(0,\tau^2\mathrm{I}) ~ ,
\end{eqnarray}
where $\phi(v;\tau^2\mathrm{I})$ is the $n$-dimensional density of $\mathcal{N}(0,\tau^2\mathrm{I})$. It can be shown that this formula is valid if $V$ is replaced by any vector of random of variables with finite higher order moments. The author in \cite{Ye98} proved that this is an unbiased estimate of $df$ as $\tau \to 0$, that is $\lim_{\tau \to 0} \mathrm{E}_Z\left[\widehat{df}(\lambda)\right] = df(\lambda)$. It can be computed by Monte-Carlo integration with $\tau$ near 0.6 as devised in \cite{Ye98}. However, both bootstrap and Ye's method, although general and can be used for any $\Psi \in \Gamma_0(\R^L)$, are computationally prohibitive. This is the main reason we will not use them here.

Zou et al. \cite{Zou07} recently studied the degrees of freedom of the Lasso\footnote{The Lasso model correspond to the case of \eqref{eq:9} where the degradation model in \eqref{eq:3} is linear and $\Psi$ is the $\ell_1$-norm.} in the framework of SURE. They showed that for any given $\lambda$ the number of nonzero coefficients in the model is an unbiased and consistent estimate of $df$. However, for their results to hold rigorously, the matrix $\Ad=\Hd\Phi$ in the Lasso must be over-determined $L < n$ with $\mathrm{Rank}(\Ad) = L$. Nonetheless, one can show that their intuitive estimator can be extended to the under-determined case (i.e. $L \geq n$) under the so-called (UC) condition of \cite{Dossal07}; see Theorem 2 in that reference. This will yield an unbiased estimator of $df$, but consistency would be much harder to prove since it requires that the Gram matrix $\Ad\Tr\Ad$ is {\textit{positive-definite}} which only happens in the special case of $\Phi$ an orthogonal basis and $\mathrm{ker}\parenth{\Hd} = \emptyset$. Furthermore, even with the $\ell_1$ norm, extending this simple estimator rigorously to our setting faces two additional serious difficulties beside underdeterminacy of $\Ad$: namely the non-linearity of the degradation equation \eqref{eq:3} and the positivity constraint in \eqref{eq:9}. 

Following this discussion, it appears clearly that estimating $df$ is either computationally intensive (bootstrap or perturbation techniques), or analytically difficult to derive. In this paper, in the same vein as \cite{Zou07}, we conjecture that a simple estimator of $df$, is given by the cardinal of the support of $\alpha^{\star}$. That is, from \eqref{eq:10}-\eqref{eq:5}
\be
\widehat{df}(\lambda) = \card\left\{i=1,\ldots,L ~ \big|~ |\alpha^{\star}_i| \geq \lambda\mu \right\} ~ .
\ee 
With such simple formula on hand, expression of the model selection criteria GCV in \eqref{eq:gcv1} is readily available.


\begin{figure}[ht]
  \centering
  \begin{tabular}{@{ }c@{ }c@{ }}
    \hspace{-1.1cm}
    \includegraphics[width=0.54\linewidth]{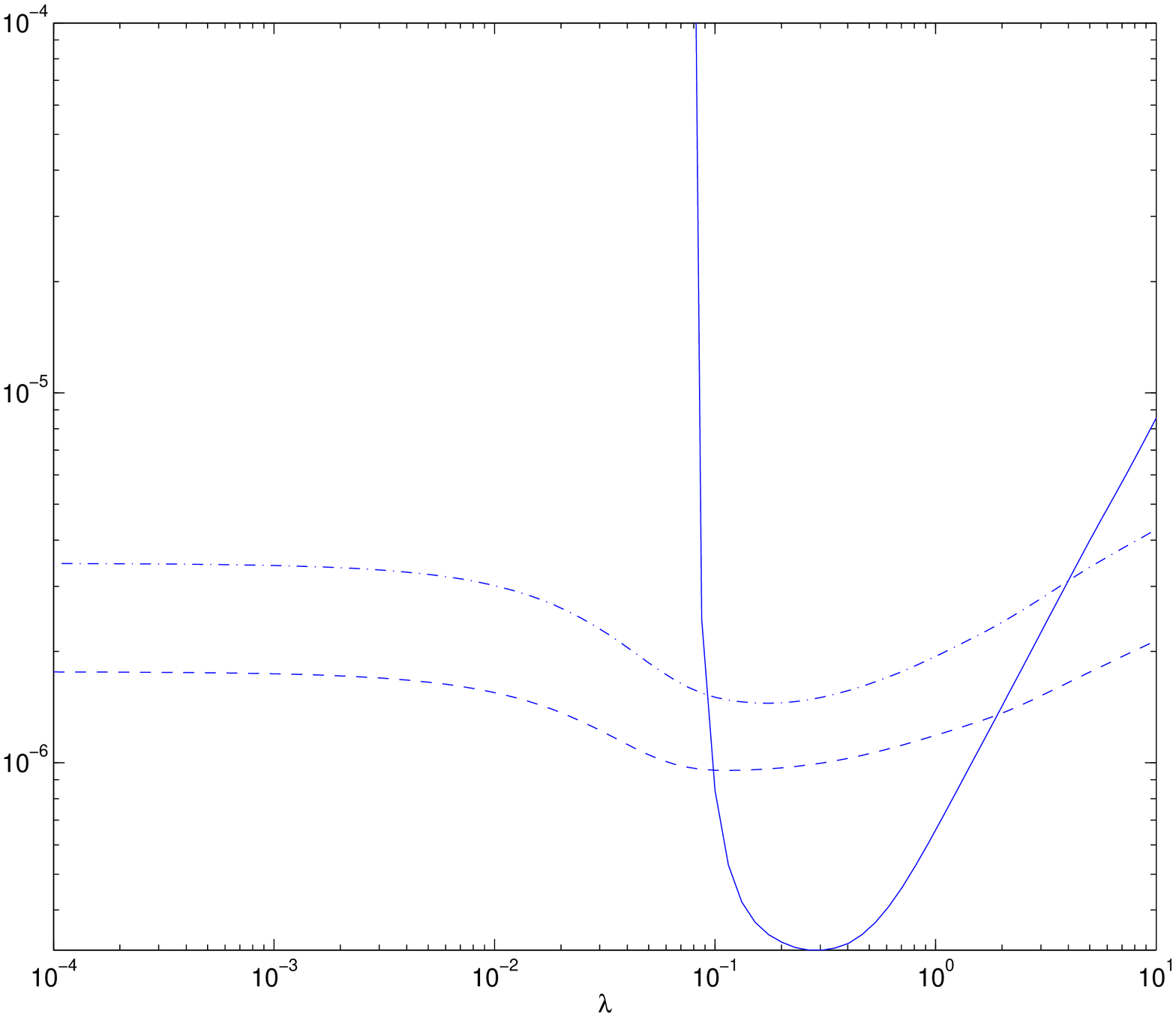} &
    \includegraphics[width=0.54\linewidth]{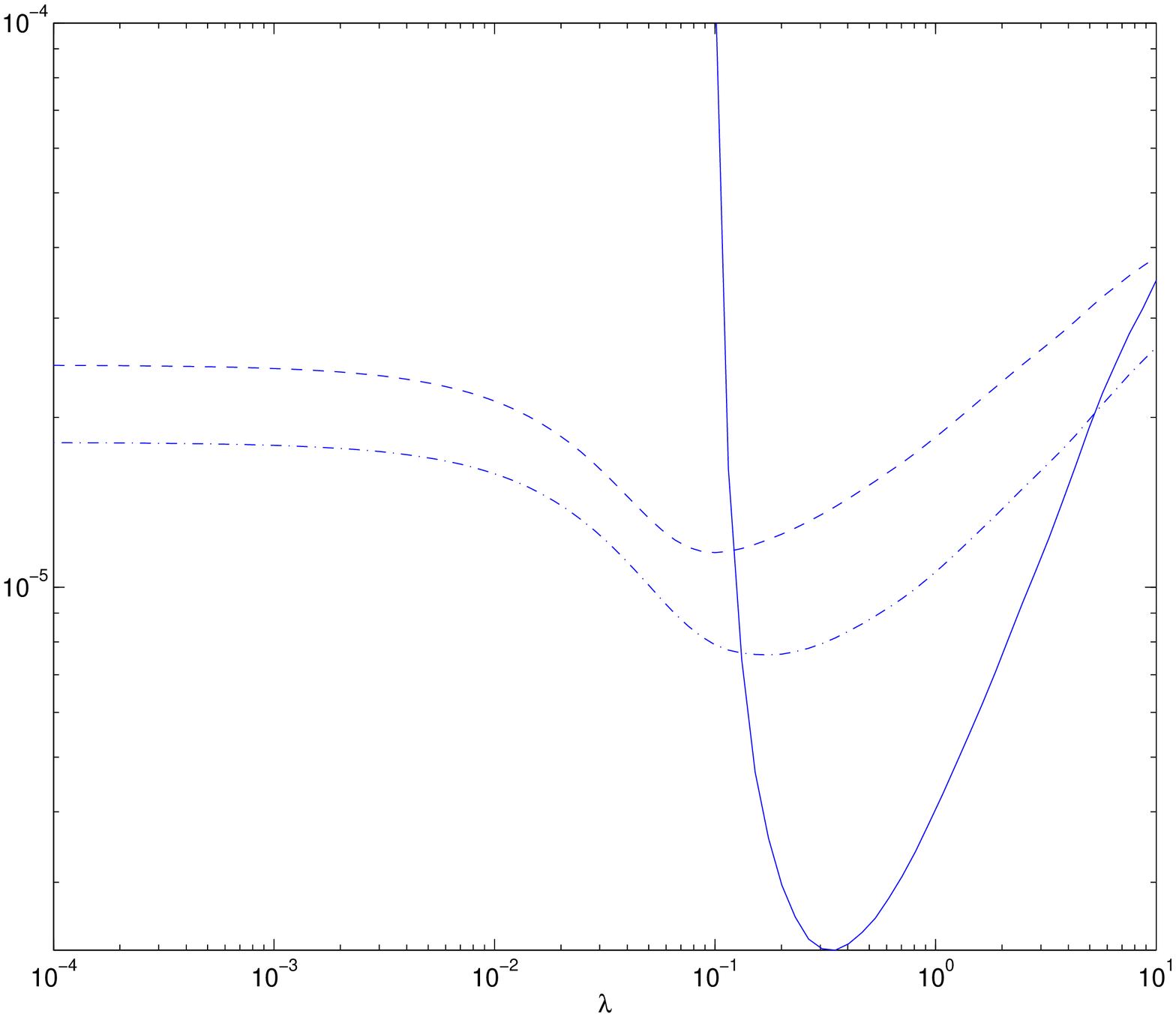} \\
    \hspace{-1.1cm} (a) &  (b)
  \end{tabular}
  \caption{GCV for the Cameraman (a) and the Neuron phantom (b). The translation-invariant discrete wavelet transform was used with the Cameraman image, and the curvelet transform with the Neuron phantom. The solid line represents the GCV, the dashed line the MSE and the dashed-dotted line the MAE.}
  \label{fig:gcv-camera-lam}
\end{figure}

Although this formula is only an approximation, in all our experiments, it performed reasonably well. This is testified by
Fig.~\ref{fig:gcv-camera-lam}(a) and (b) which respectively show the behavior of the GCV
as a function of $\lambda$ for two images: the Cameraman portrayed in
Fig.~\ref{fig:cameraman}(a) and the Neuron phantom shown in Fig.~\ref{fig:neuron}(a). As
the ground-truth is known in the simulation, we computed for each $\lambda$ the mean
absolute-error (MAE){\textemdash}well adapted to Poisson noise as it is closely related to the squared Hellinger distance \cite{Barron91}{\textemdash} as well as the mean square-error (MSE) between the deconvolved and true image. We can clearly see that the GCV reaches its minimum close to those of the MAE and the MSE. Even though the regularization parameter dictated by the GCV criterion is slightly higher than that of the MSE, which may lead to a somewhat over-smooth estimate.

\subsection{Computational complexity and implementation details}
\label{sec:comp-compl-impl}

The bulk of computation of our deconvolution algorithm is invested in applying $\Phi$
(resp. $\Hd$) and its adjoint $\Phi\Tr$ (resp. $\Hd\Tr$). These operators are never
constructed explicitly, rather they are implemented as fast implicit operators taking a
vector $x$, and returning $\Phi x$ (resp. $\Phi\Tr x$) and $\Hd x$ (resp. $\Hd\Tr
x$). Multiplication by $\Hd$ or $\Hd\Tr$ costs two FFTs, that is $2n \log n$ operations
($n$ denotes the number of pixels). The complexity of $\Phi$ and $\Phi\Tr$ depends on the
transforms in the dictionary: for example, the orthogonal wavelet transform costs
$\mathcal{O}(n)$ operations, the translation-invariant discrete wavelet transform (TI-DWT) costs $\mathcal{O}(n\log n)$, 
the curvelet transform costs $\mathcal{O}(n\log n)$, etc. Let
$V_\Phi$ denote the complexity of applying the analysis or synthesis operator. Define
$N_{\mathrm{FB}}$ and $N_{\mathrm{DR}}$ as the number of iterations in the
forward-backward algorithm and the Douglas-Rachford sub-iteration, and recall that $L$ is
the number of coefficients. The computational complexities of our iterations \eqref{eq:5}
and \eqref{eq:tseng} are summarized below:

\begin{center}
  \begin{tabular}{|c|c|c|} \hline Algorithm & \multicolumn{2}{|c|}{Computational complexity bounds} \\
    \hline\hline & $\Phi$ orthobasis & $\Phi$ tight frame \\
    \hline \eqref{eq:5} & $N_{\mathrm{FB}}\parenth{4n\log n + N_{\mathrm{DR}}\parenth{2V_\Phi + \mathcal{O}(n)}}$ &
    $N_{\mathrm{FB}}\parenth{4n\log n + 2V_\Phi + N_{\mathrm{DR}}(2V_\Phi + \mathcal{O}(L))}$ \\
    \hline \eqref{eq:tseng} & $N_{\mathrm{FB}}\parenth{8n\log n + 2V_\Phi + \mathcal{O}(n)}$ &
    $N_{\mathrm{FB}}\parenth{8n\log n + 6 V_\Phi + \mathcal{O}(L)}$ \\\hline
  \end{tabular}
\end{center}

The orthobasis case requires less multiplications by $\Phi$ and $\Phi\Tr$ because in that
case, $\Phi$ is a bijective linear operator. Thus, the optimization problem \eqref{eq:9}
can be equivalently written in terms of image samples instead of coefficients, hence
reducing computations in the corresponding iterations \eqref{eq:5} and \eqref{eq:tseng}.

For our implementation, as in Algorithm~\ref{algo:1}, we have taken $a_t = b_t \equiv 0$ and $\beta_t \equiv 1$ in \eqref{eq:5}, 
and $\nu_t \equiv 1/2$ in \eqref{eq:proxtframe1}. As the PSF $h$ in our experiments is low-pass normalized to a unit
sum, $\norm{\Hd}_2^2 = 1$. $\Psi$ was the $\ell_1$-norm, leading to
soft-thresholding. Furthermore, in order to accelerate the computation of $\prox_{f_2}$ in
\eqref{eq:5}, the Douglas-Rachford sub-iteration \eqref{eq:proxtframe1} was only run once
(i.e. $N_{\mathrm{DR}}=1$) starting with $\gamma^0 = \va$. In this case, one can check that if $\gamma^0 \in \C'$, then this leads
to the "natural" formula :
\begin{equation*}
  \prox_{\reg} (\va) = \Prj_{\C'} \circ \prox_{\tfrac{\lambda}{2}\Psi} (\va).
\end{equation*}

In our experimental studies, the GCV-based selection of $\lambda$ was run using the
forward-backward algorithm \eqref{eq:5} which has a lower computational burden than
\eqref{eq:tseng} (see above table for computational complexities). Once $\lambda$ was
objectively chosen by the GCV procedure, the deconvolution algorithm was applied using
\eqref{eq:tseng} to exempt the user from the choice of the relaxation parameter $\mu$.

\section{Results}
\label{sec:results}

\subsection{Simulated data}
\label{sec:simulated-data}

The performance of our approach has been assessed on several test images: a $128 \times
128$ neuron phantom \cite{WillettURL}, a $370 \times 370$ confocal microscopy image of
micro-vessel cells \cite{ImageJ}, the Cameraman ($256 \times 256$), a $512 \times 512$
simulated astronomical image of the Hubble Space Telescope Wide Field Camera of a distant
cluster of galaxies \cite{Starck2006}.  Our algorithm was compared to RL with total
variation regularization (RL-TV \cite{Dey2004}), RL with multi-resolution support wavelet
regularization (RL-MRS \cite{Starck95}), fast translation invariant tree-pruning
reconstruction combined with an EM algorithm (FTITPR \cite{Willett2004}) and the naive
proximal method that would treat the noise as if it were Gaussian (NaiveGauss
\cite{Vonesch2007}). For all results presented, each algorithm was run with $N_\mathrm{FB}=200$
iterations, enough to reach convergence. For all results below, $\lambda$ was selected
using the GCV criterion for our algorithm. For fair comparison to \cite{Vonesch2007},
$\lambda$ was also chosen by adapting our GCV formula to the Gaussian noise.

\begin{figure}[ht]
  \centering
  \begin{tabular}{@{ }c@{ }c@{ }c@{ }}
    \includegraphics[width=0.3\linewidth]{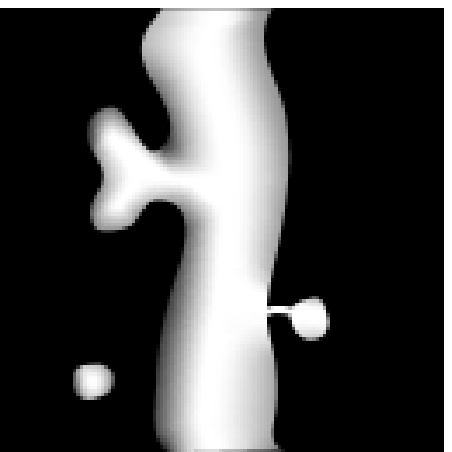} &
    \includegraphics[width=0.3\linewidth]{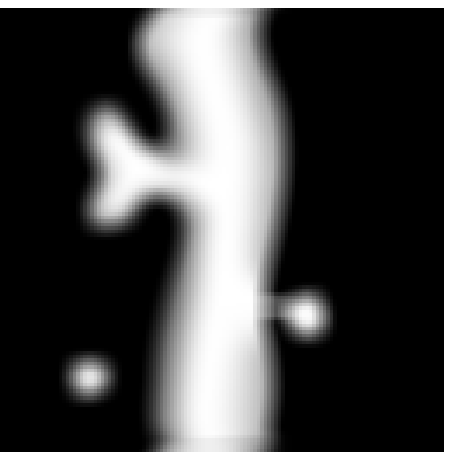} &
    \includegraphics[width=0.3\linewidth]{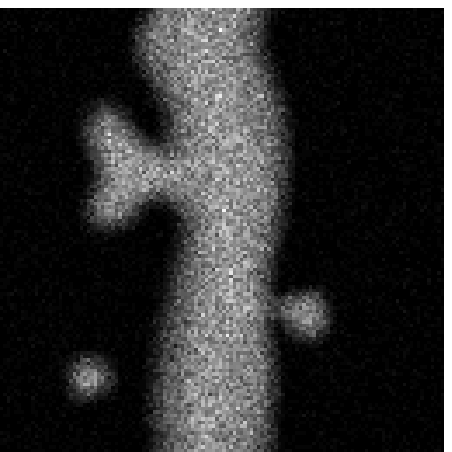} \\
    (a) & (b) & (c)	\\
    \includegraphics[width=0.3\linewidth]{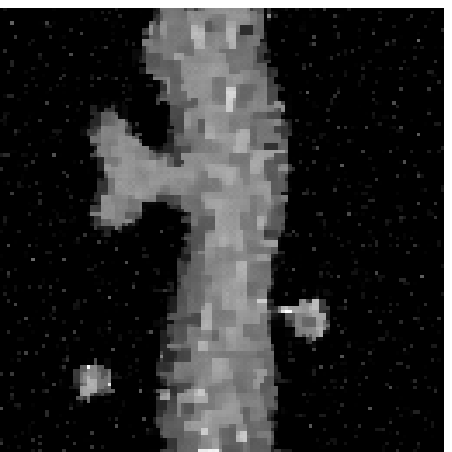} &
    \includegraphics[width=0.3\linewidth]{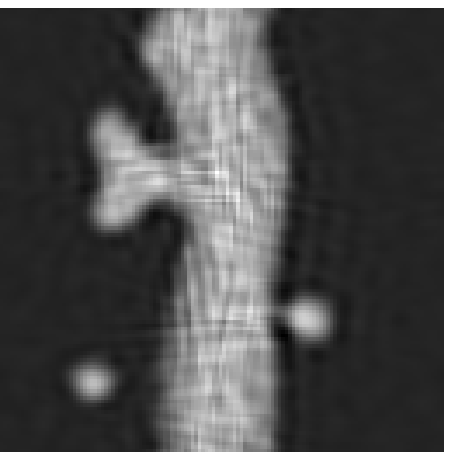} &
    \includegraphics[width=0.3\linewidth]{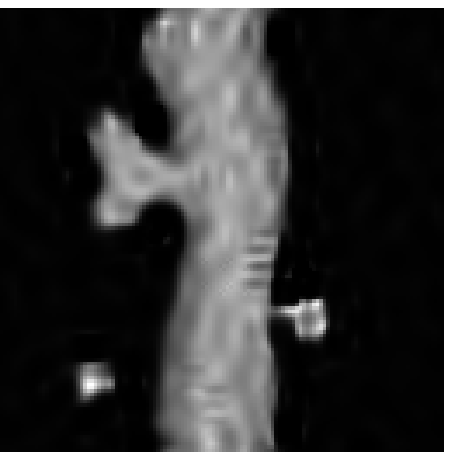} \\
    (d) & (e) & (f) \\
    \includegraphics[width=0.3\linewidth]{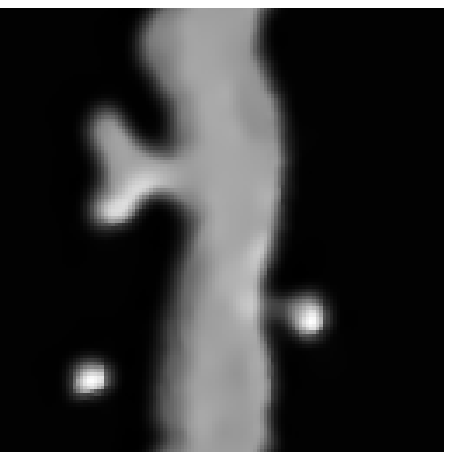}&
    \includegraphics[width=0.3\linewidth]{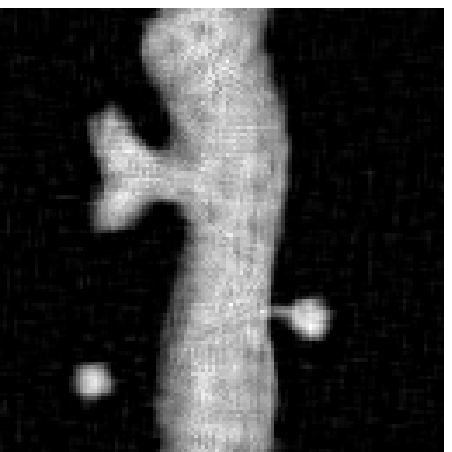}& \\
    (g) & (h) &     \\
  \end{tabular}
\caption{Deconvolution of a simulated neuron (Intensity $\leq$ 30). (a)
    Original, (b) Blurred, (c) Blurred\&noisy, (d) RL-TV \cite{Dey2004}, (e) NaiveGauss
    \cite{Vonesch2007}, (f) RL-MRS \cite{Starck2006}, (g) FTITPR \cite{Willett2004}, (h)
    Our Algorithm.}
  \label{fig:neuron}
\end{figure}

Fig.\ref{fig:neuron}(a), depicts a phantom of a neuron with a mushroom-shaped spine. The
maximum intensity is 30. Its blurred (using a $7\times 7$ moving average) and blurred+noisy
versions are in (b) and (c). With this neuron, and for NaiveGauss and our approach, the
dictionary $\Phi$ contained the curvelet tight frame \cite{CandesFDCT05}. The
deconvolution results are shown in Fig.\ref{fig:neuron}(d)-(h). As expected at this
intensity level, the NaiveGauss algorithm performs quite badly, as it does not fit
the noise model at this intensity regime. It turns out that NaiveGauss under-regularizes
the estimate and the Poisson signal-dependent noise is not always under control. This behavior of
NaiveGauss, which was predictable at this intensity level, will be observed on almost all tested
images. RL-TV does a good job at deconvolution but the background is dominated by
artifacts, and the restored neuron has staircase-like artifacts typical of TV
regularization. Our approach provides a visually pleasant deconvolution result on this
example. It efficiently restores the spine, although the background is not fully
cleaned. RL-MRS also exhibits good deconvolution results. On this image, FTITPR provides a
well smoothed estimate but with almost no deconvolution.

\begin{figure}[ht]
  \hspace*{-1cm}
  \begin{tabular}{@{}c@{}@{}c@{}@{}c@{}}
  \includegraphics[width=0.37\linewidth]{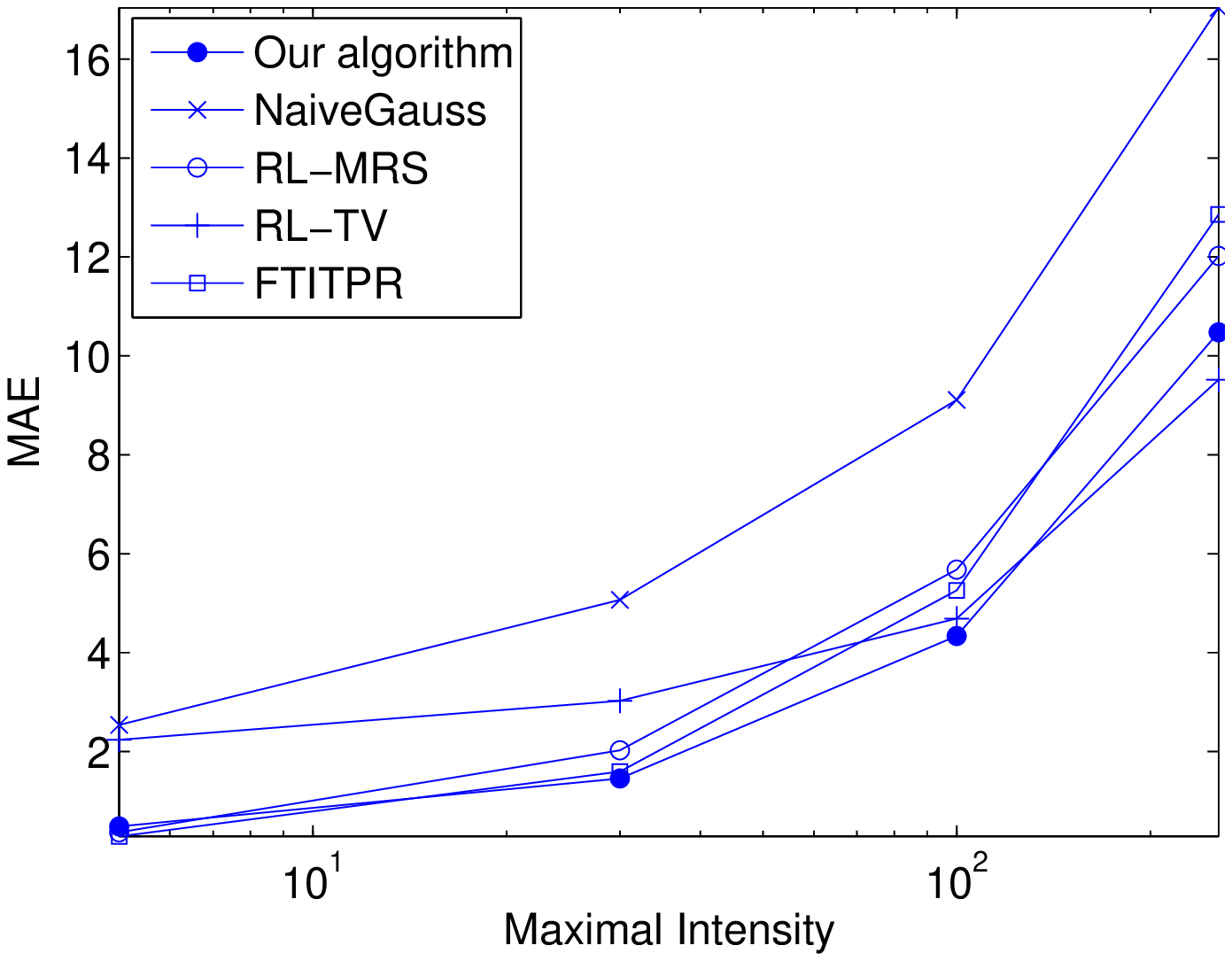} &
  \includegraphics[width=0.37\linewidth]{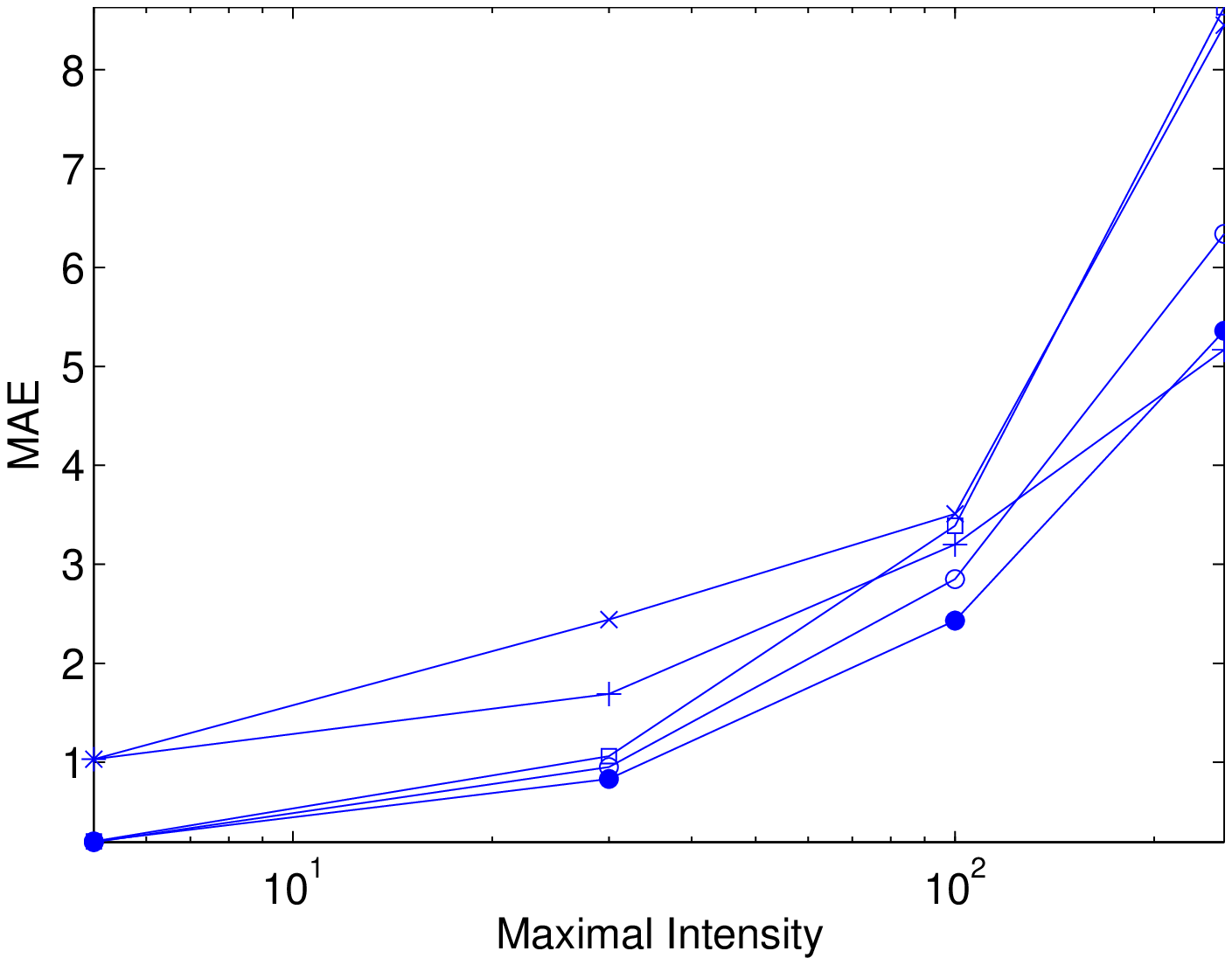} &
  \includegraphics[width=0.37\linewidth]{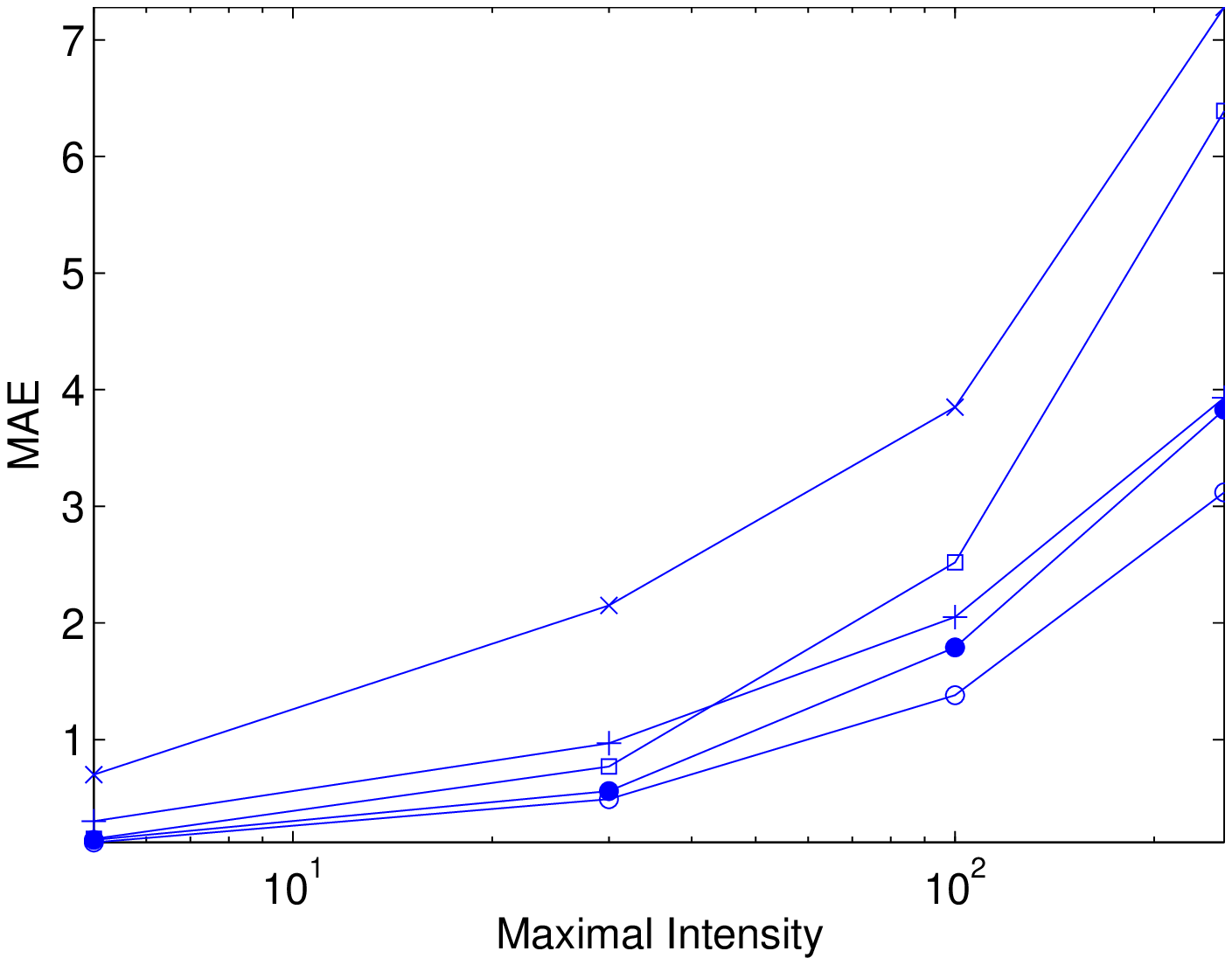} \\
  (a) & (b) & (c)
  \end{tabular}
  \caption{Average MAE of all algorithms as a function of the intensity level. (a) Cameraman, (b) Neuron phantom, (c) Cell.}
  \label{fig:intens}
\end{figure}

These qualitative visual results are confirmed by quantitative measures of the quality of
deconvolution, where we used both the MAE and the traditional MSE criteria. 
At each intensity value, 10 noisy and blurred replications were generated
and and the MAE was computed for each deconvolution algorithm. The average MAE over the 10
replications are given in Fig.~\ref{fig:intens} (similar results were obtained for the
MSE, not shown here). In general, our algorithm performs very well at all intensity regimes (especially at medium to low).
The NaiveGauss is among the worst algorithms at low intensity levels. Its performance becomes better as the intensity increases which was expected. RL-MRS is effective at low and medium intensity levels and is even better than our algorithm on the Cell image. 
RL-TV underperforms all algorithms at low intensity. We suspect the staircase-like artifacts of TV-regularization to be responsible
for this behavior. At high intensity, RL-TV becomes competitive and its MAE comparable to ours.

\begin{figure}[ht]
  \centering
    \begin{tabular}{@{ }c@{ }c@{ }c@{ }}
      \includegraphics[width=0.3\linewidth]{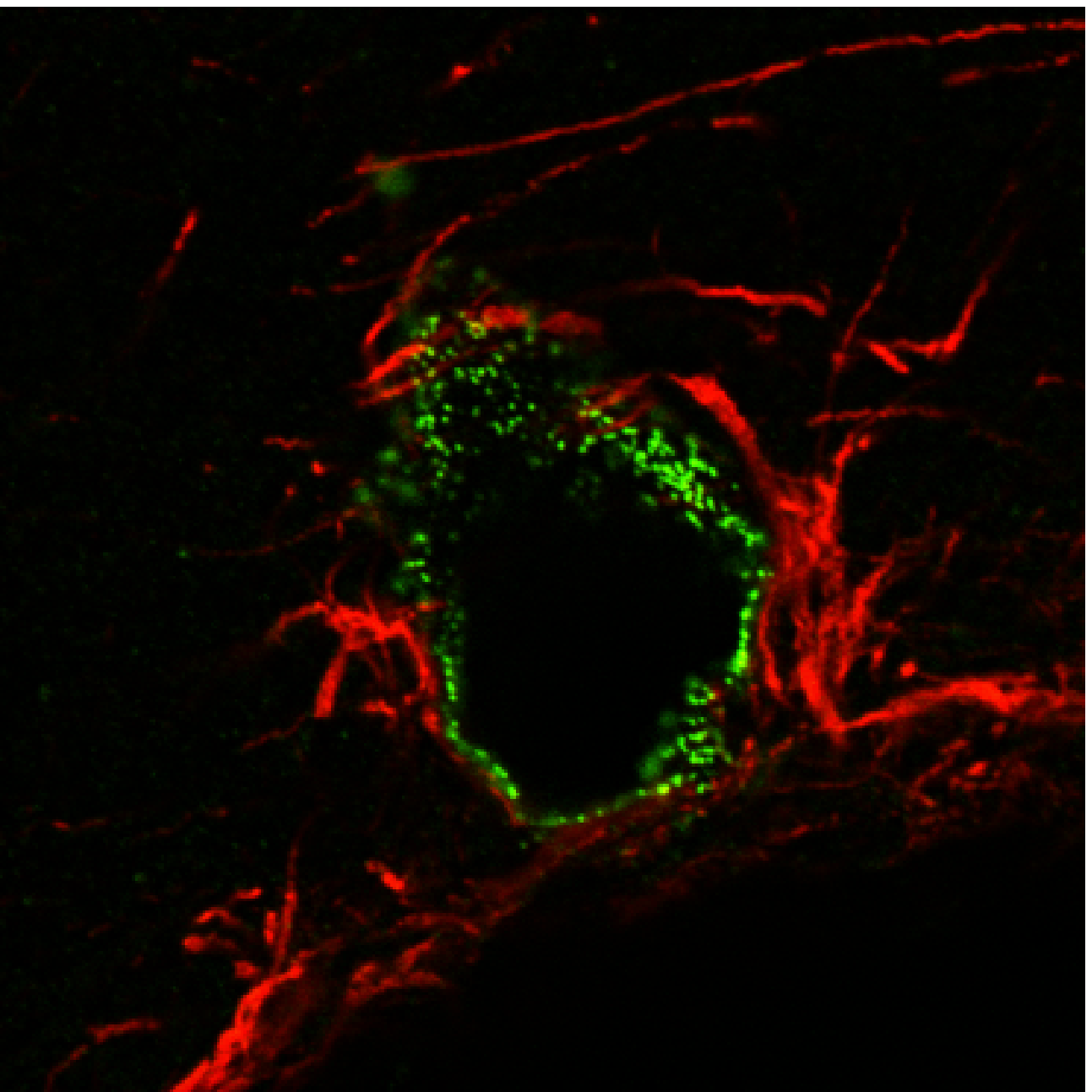} &
      \includegraphics[width=0.3\linewidth]{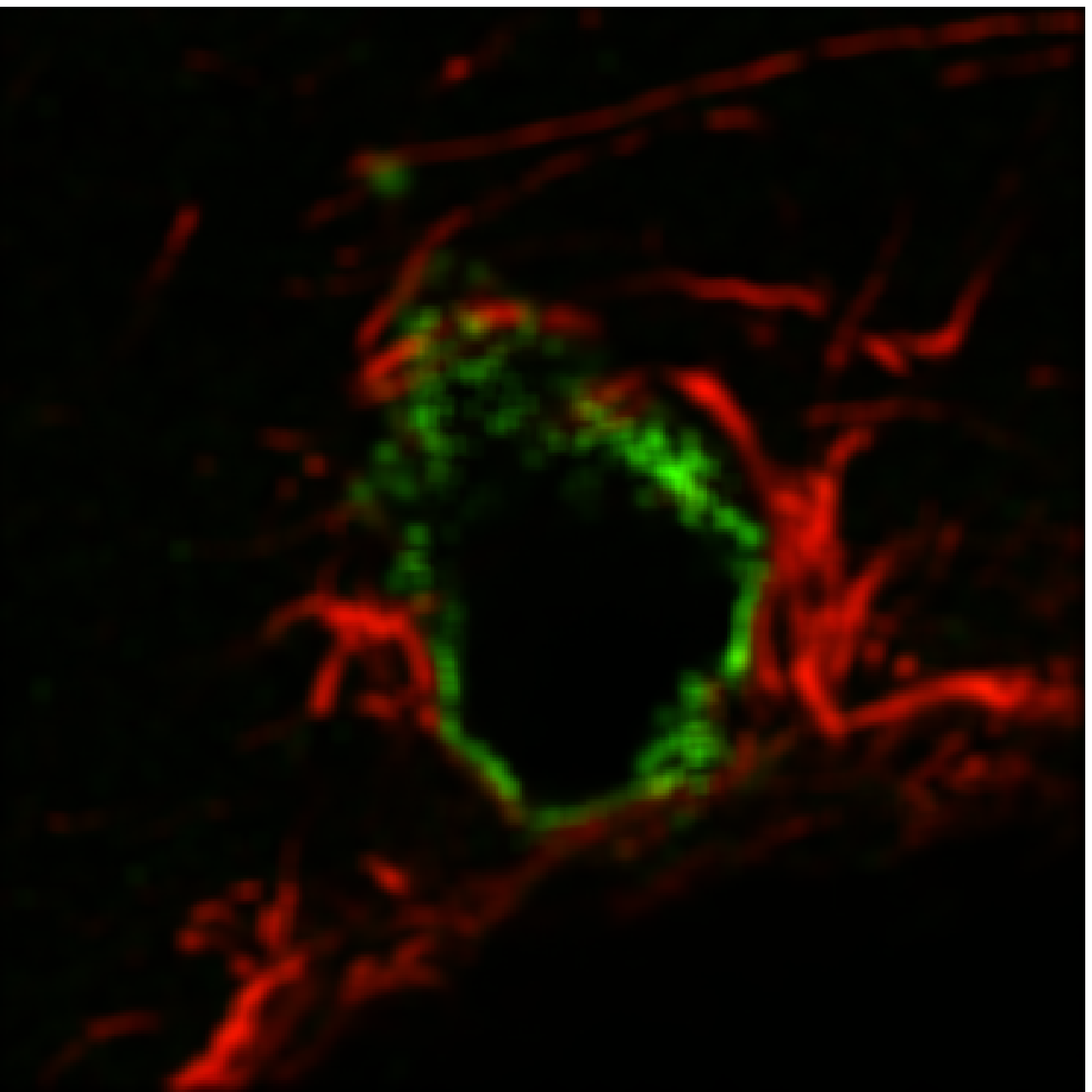} &
      \includegraphics[width=0.3\linewidth]{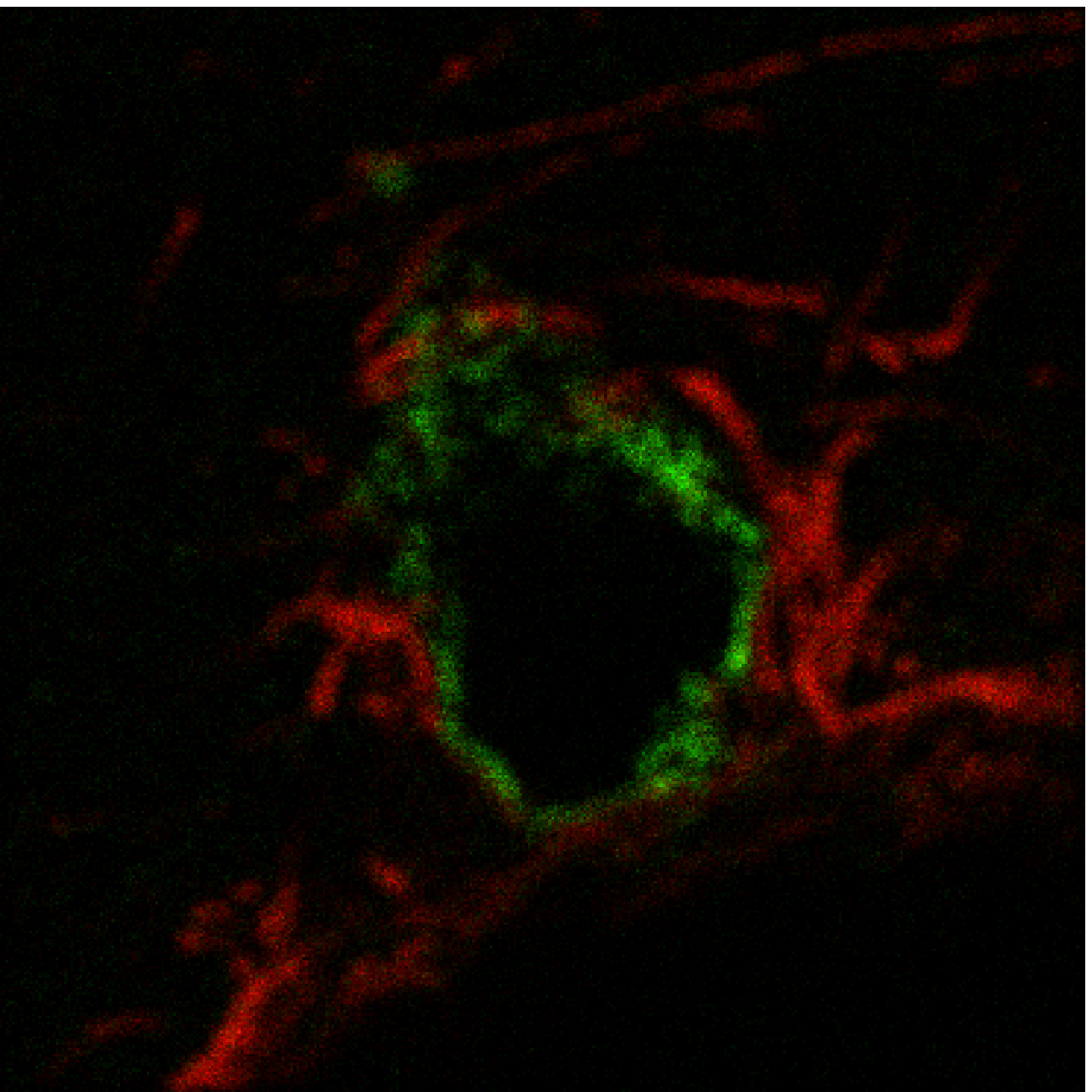} \\
      (a) & (b) & (c)	\\
      \includegraphics[width=0.3\linewidth]{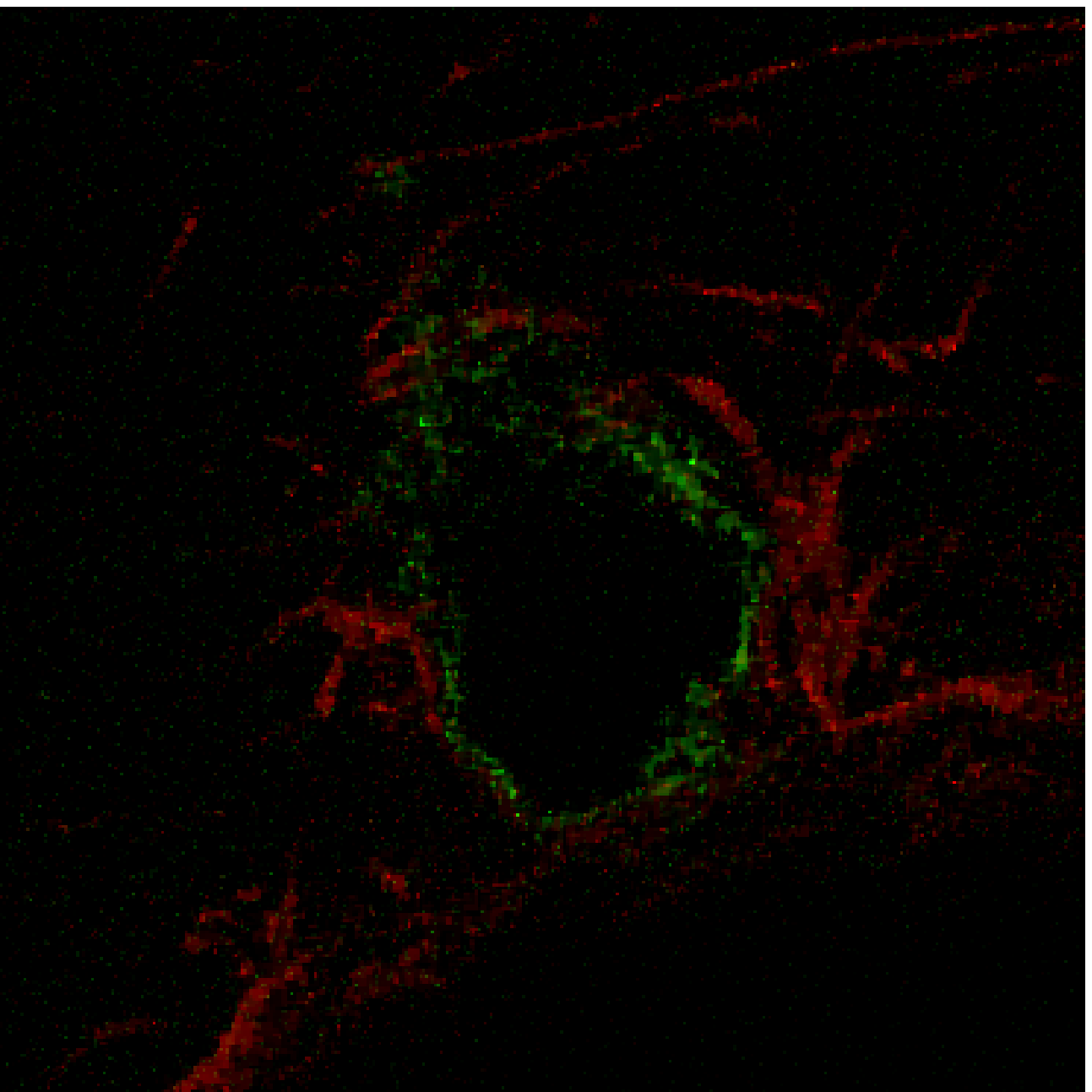} &
      \includegraphics[width=0.3\linewidth]{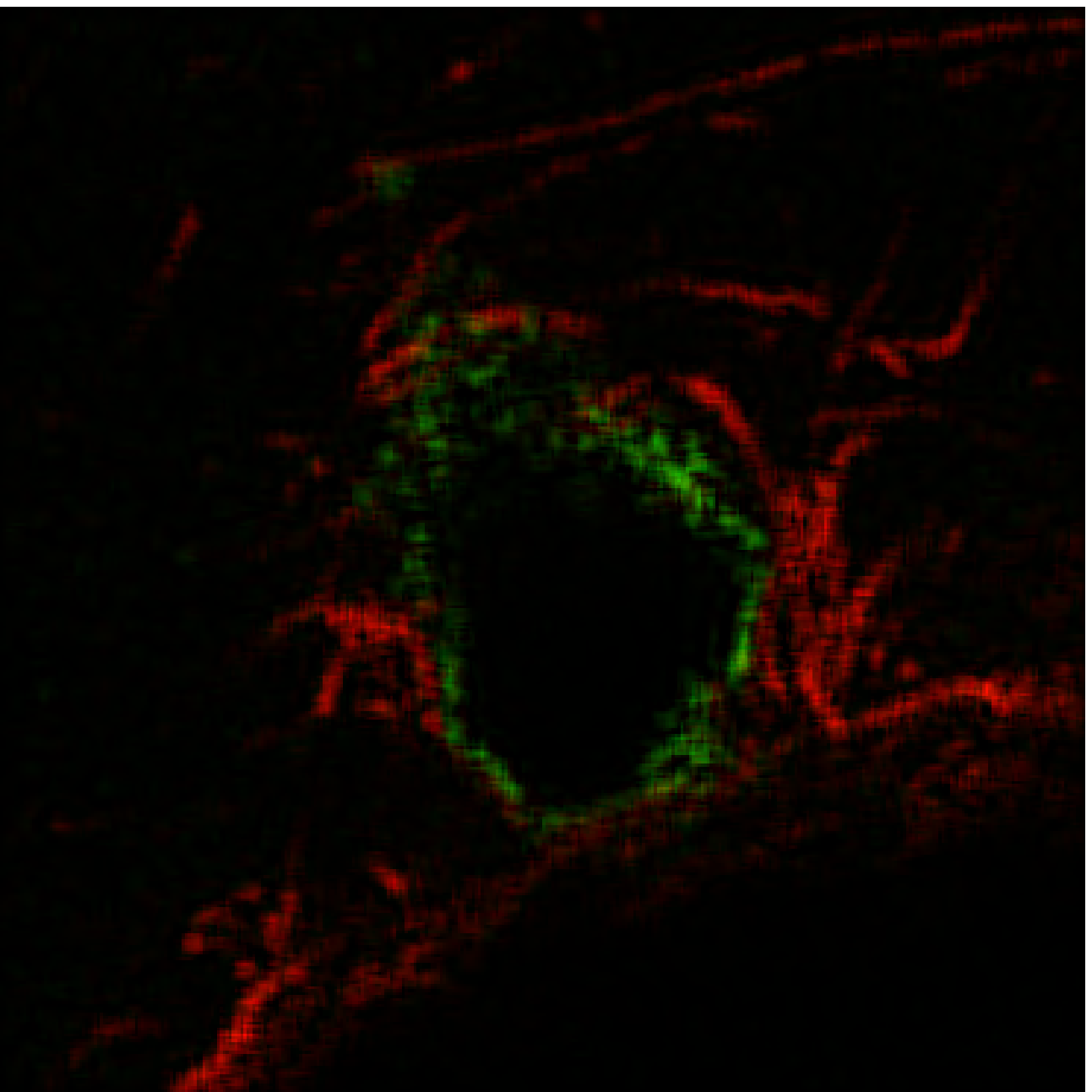} &
      \includegraphics[width=0.3\linewidth]{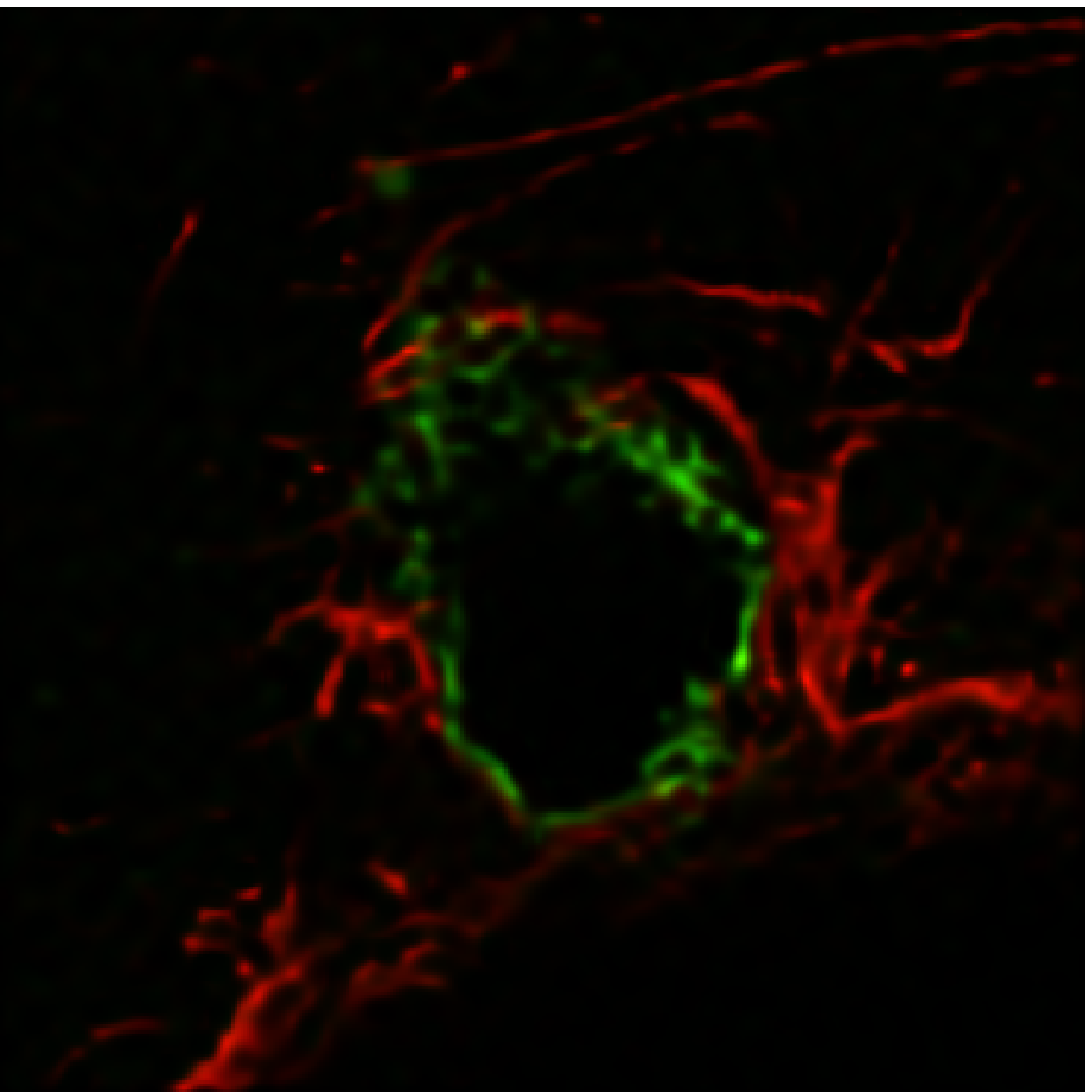} \\
      (d) & (e) & (f) \\
      \includegraphics[width=0.3\linewidth]{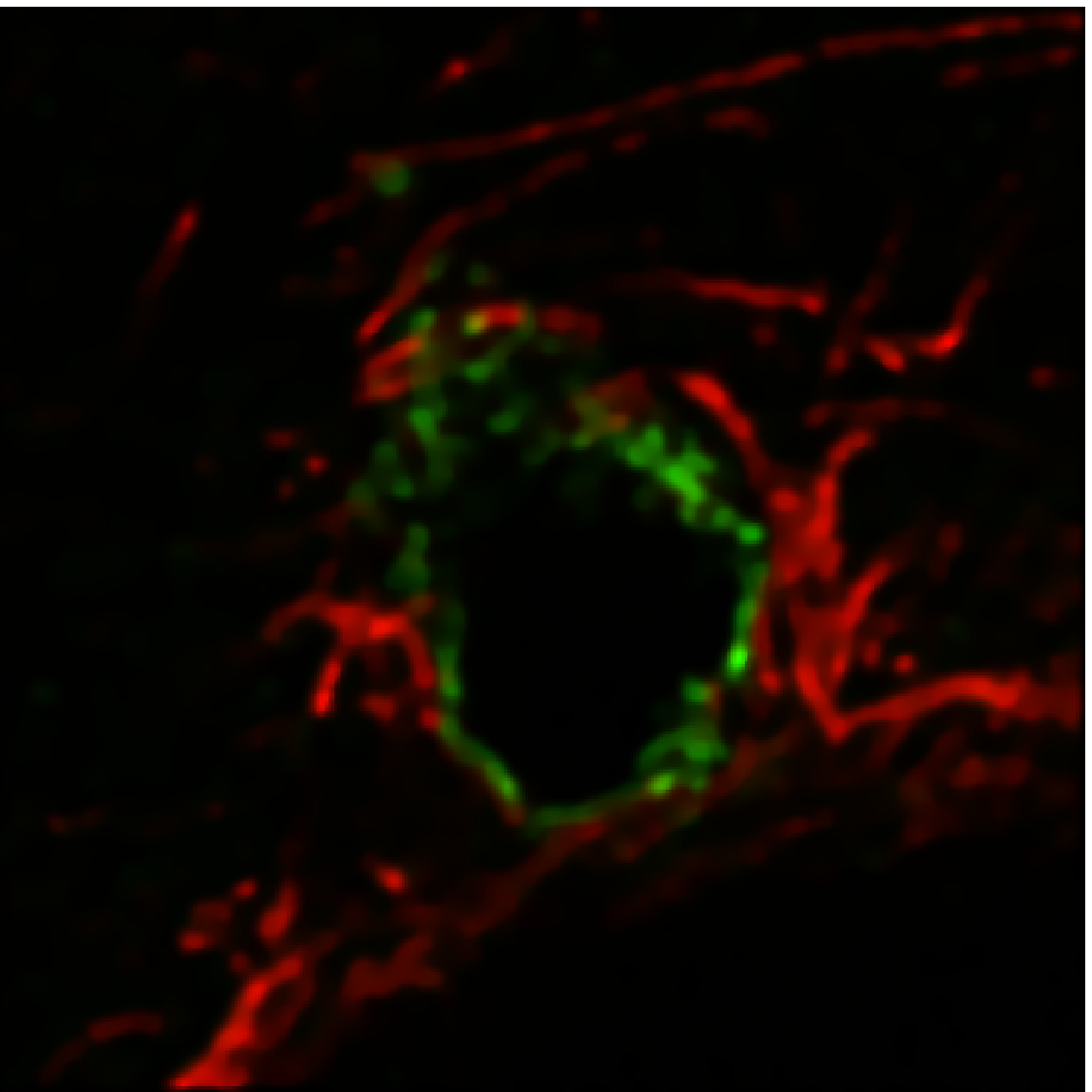} &
      \includegraphics[width=0.3\linewidth]{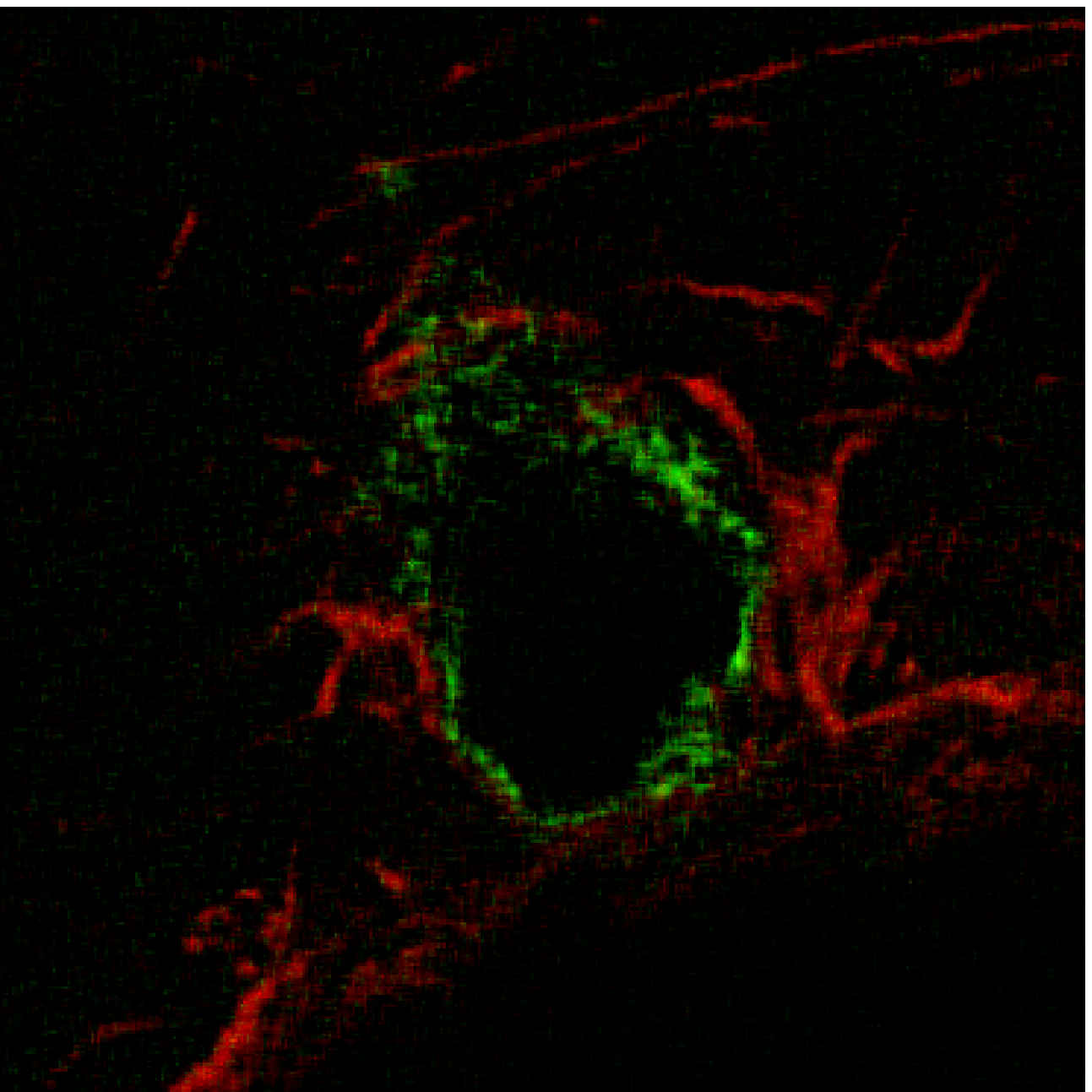} & \\
      (g) & (h) &     \\
    \end{tabular}
  \caption{{Deconvolution of a microscopy cell image (Intensity $\leq$ 30). (a) Original, (b)
      Blurred, (c) Blurred\&noisy, (d) RL-TV \cite{Dey2004}, (e) NaiveGauss
      \cite{Vonesch2007}, (f) RL-MRS \cite{Starck2006}, (g) FTITPR \cite{Willett2004} (h)
      Our Algorithm.}}
  \label{fig:cell}
\end{figure}

The same experiment as above was carried out with the confocal microscopy cell image; see
Fig.~\ref{fig:cell}. In this experiment, the PSF was a $7 \times 7$ moving average. For
the NaiveGauss and our approach, the dictionary $\Phi$ contained the TI-DWT. 
NaiveGauss deconvolution result is spoiled by artifacts. RL-TV produces a good restoration of small isolated details but with a
dominating staircase-like artifacts. FTITPR yields a somewhat oversmooth estimate, whereas 
our approach provides a sharper deconvolution result. This visual
inspection is in agreement with the MAE measures of Fig.~\ref{fig:intens}. In particular,
one can notice that RL-MRS shows the best behavior, and the performance of our approach compared to the other methods on this cell image is roughly the same as on the previous neuron image.

\begin{figure}[ht]
  \centering
    \begin{tabular}{@{ }c@{ }c@{ }c@{ }}
      \includegraphics[width=0.3\linewidth]{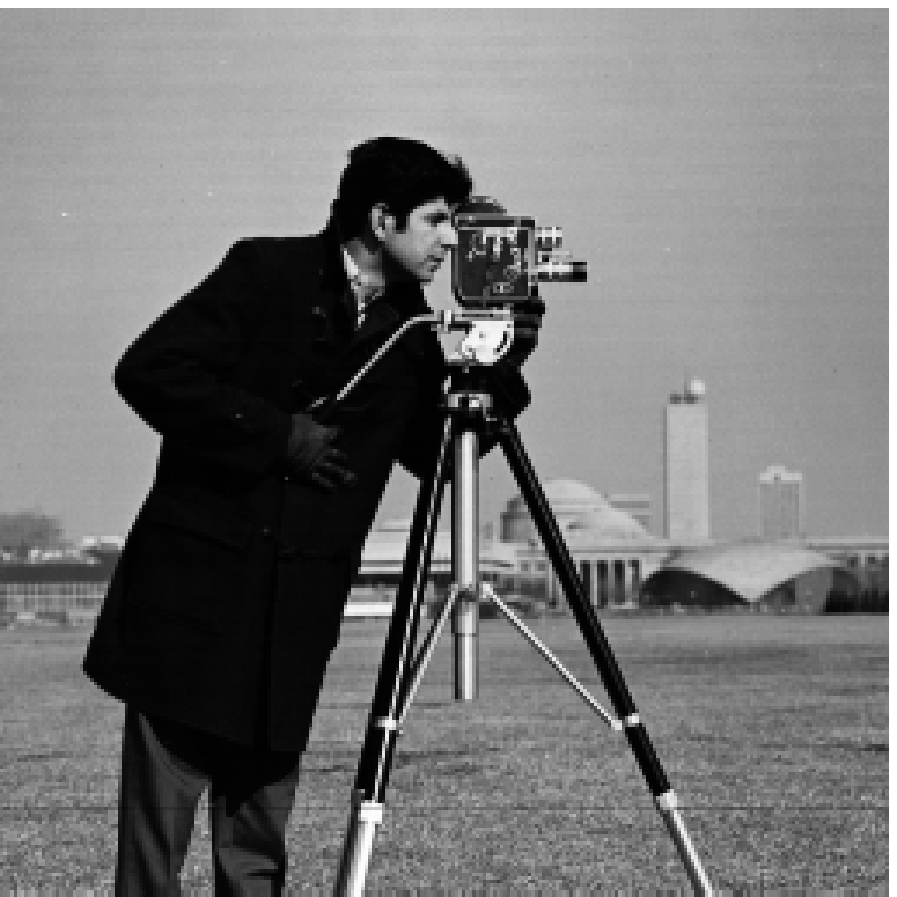} &
      \includegraphics[width=0.3\linewidth]{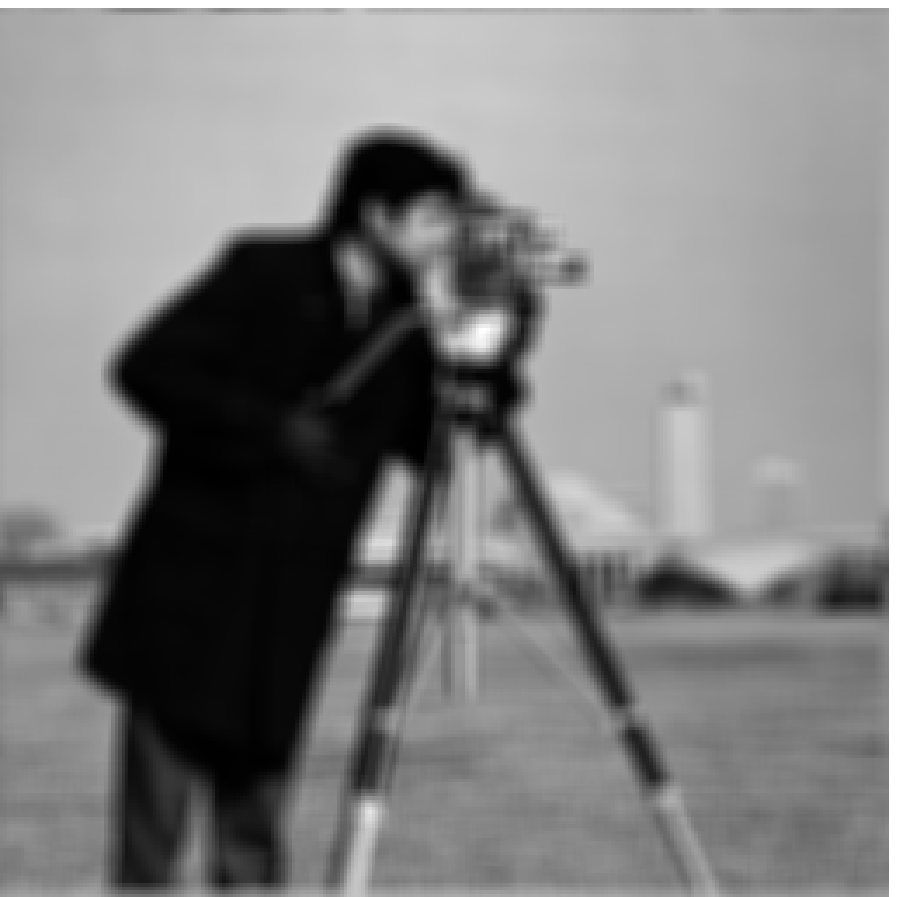} &
      \includegraphics[width=0.3\linewidth]{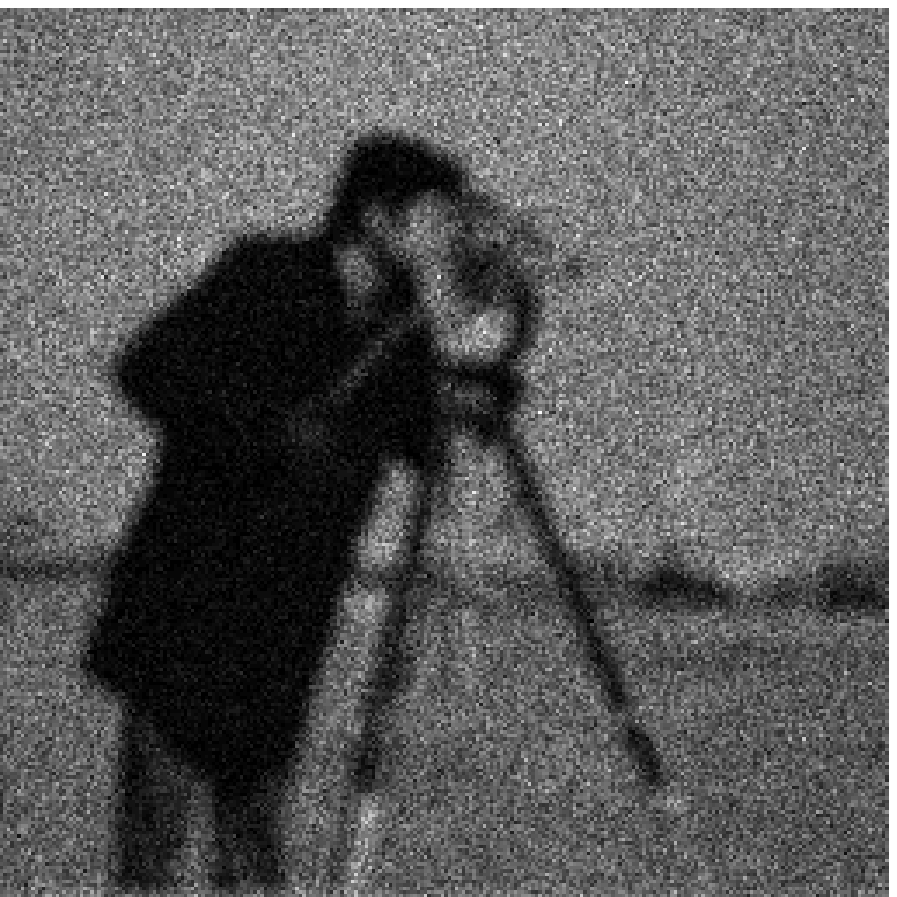} \\
      (a) & (b) & (c)	\\
      \includegraphics[width=0.3\linewidth]{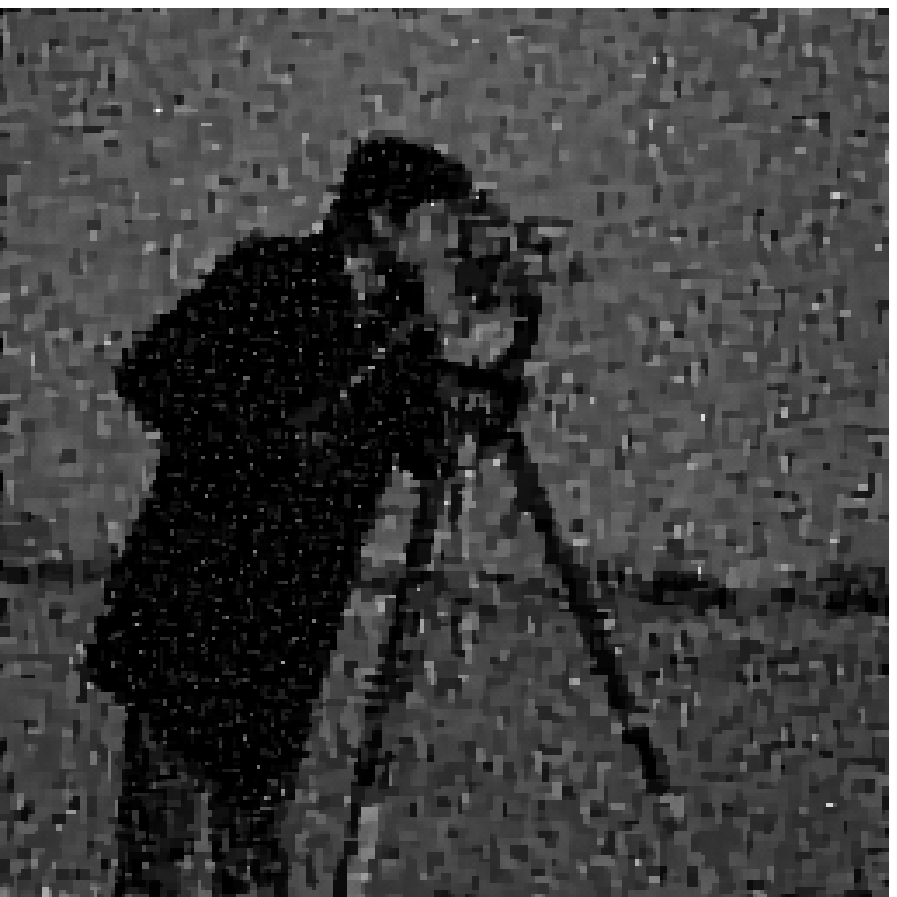} &
      \includegraphics[width=0.3\linewidth]{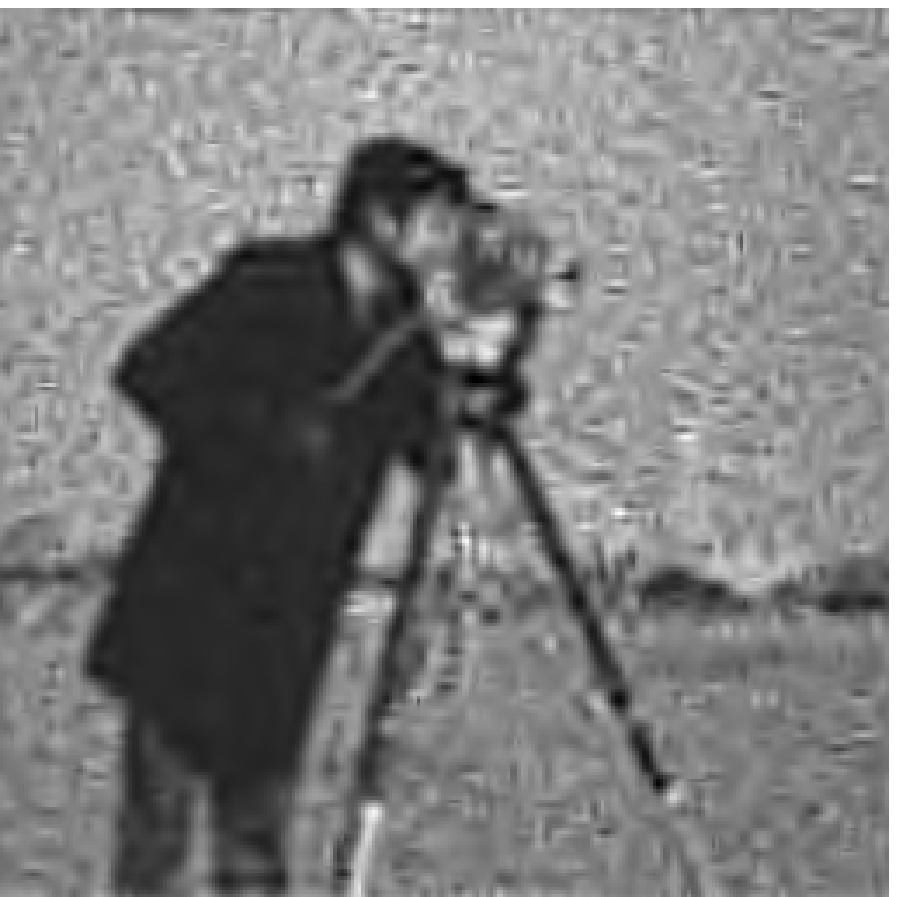} &
      \includegraphics[width=0.3\linewidth]{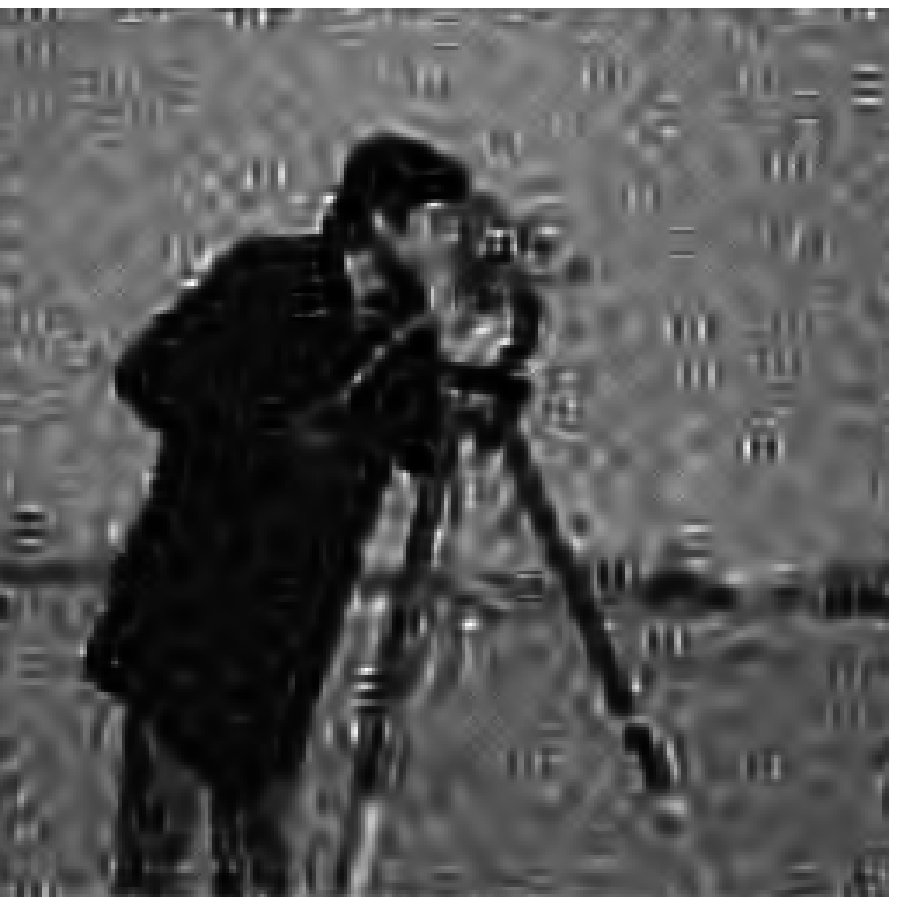} \\
      (d) & (e) & (f) \\
      \includegraphics[width=0.3\linewidth]{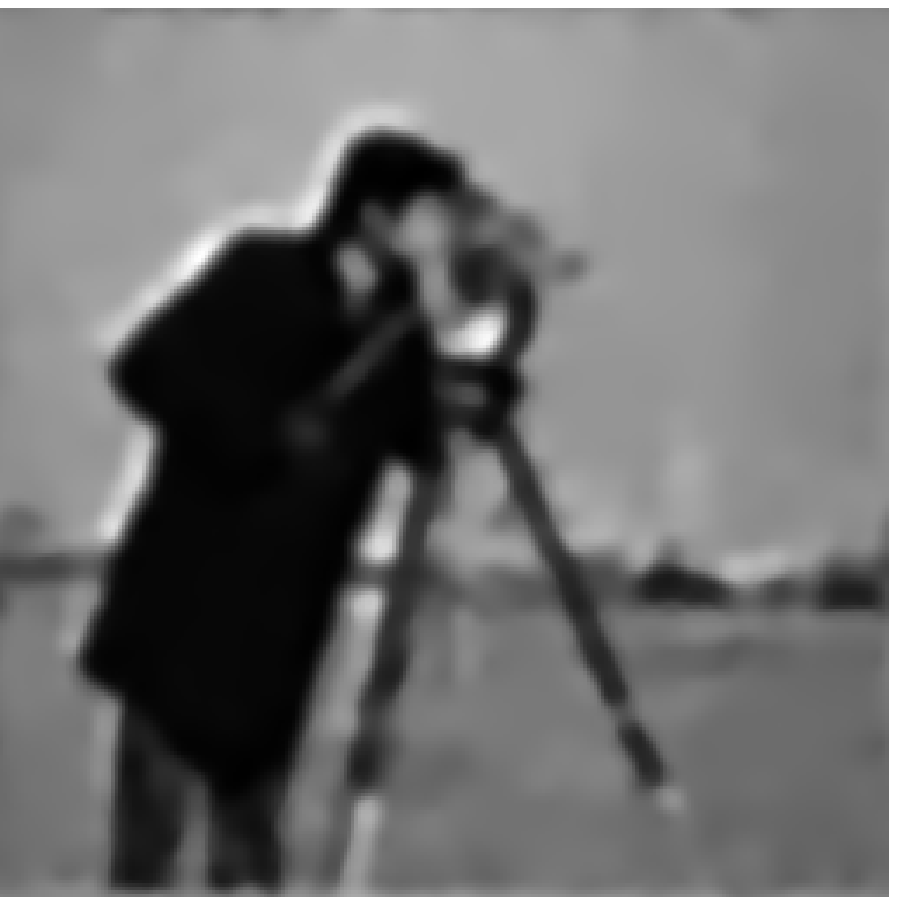} &
      \includegraphics[width=0.3\linewidth]{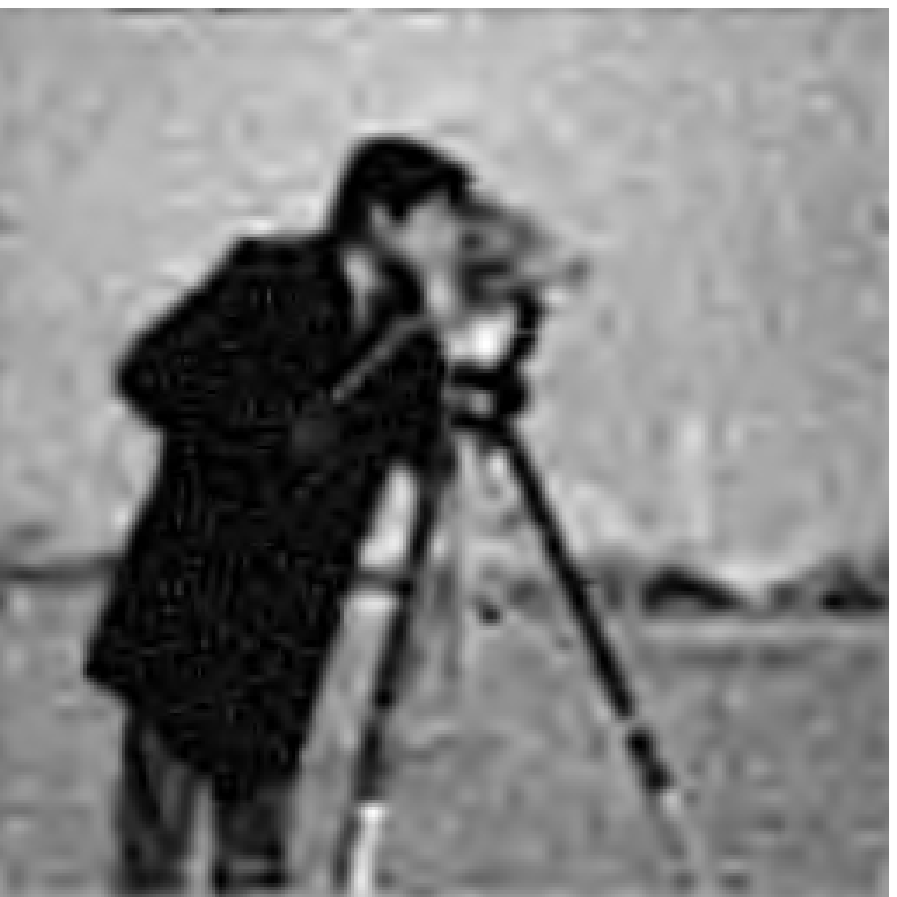}& \\
      (g) & (h) &     \\
    \end{tabular}
  \caption{{Deconvolution of the cameraman (Intensity $\leq$ 30). (a)
      Original, (b) Blurred, (c) Blurred\&noisy, (d) RL-TV \cite{Dey2004}, (e) NaiveGauss
      \cite{Vonesch2007}, (f) RL-MRS \cite{Starck2006}, (g) FTITPR \cite{Willett2004}, (h)
      Our Algorithm.}}
  \label{fig:cameraman}
\end{figure}

Fig.\ref{fig:cameraman}(a) depicts the result of the experiment on the Cameraman with
maximum intensity of 30. The PSF was the same as above. Again, the dictionary contained the
TI-DWT frame. One may notice that the degradation in Fig.\ref{fig:cameraman}(c) is quite
severe. Our algorithm provides the most visually pleasing result with a good balance
between regularization and deconvolution, although some artifacts are persisting. RL-MRS
manages to deconvolve the image with more artifacts than our approach, and suffers from a
loss of photometry. Again, FTITPR gives an oversmooth estimate with many missing
details. Both RL-TV and NaiveGauss yield results with many artifacts. This
visual impression is in agreement with the MAE values in Fig.~\ref{fig:intens}.

To assess the computational cost of the compared algorithms, Tab.~\ref{tab:time}
summarizes the execution times on the Cameraman image with an Intel PC Core 2 Duo
2GHz, 2Gb RAM. Except RL-MRS which is written in C++, all other algorithms were
implemented in Matlab.

\begin{table}[ht]
  \centering
  \footnotesize{
    \begin{tabular}{|c|c|}
      \hline Method		        &  Time (in s)\\
      \hline Our method			&  88         \\
      \hline NaiveGauss			&  71         \\
      \hline RL-MRS                     &  99.5       \\
      \hline RL-TV                      &  15.5       \\
      \hline
    \end{tabular}}
  \caption{{Execution times for the simulated $256 \times 256$ Cameraman image using the TI-DWT ($N_\mathrm{FB} = 200$).}}
  \label{tab:time}
\end{table}

\begin{figure}[ht]
  \centering
    \begin{tabular}{@{ }c@{ }c@{ }c@{ }}
      \includegraphics[width=0.3\linewidth]{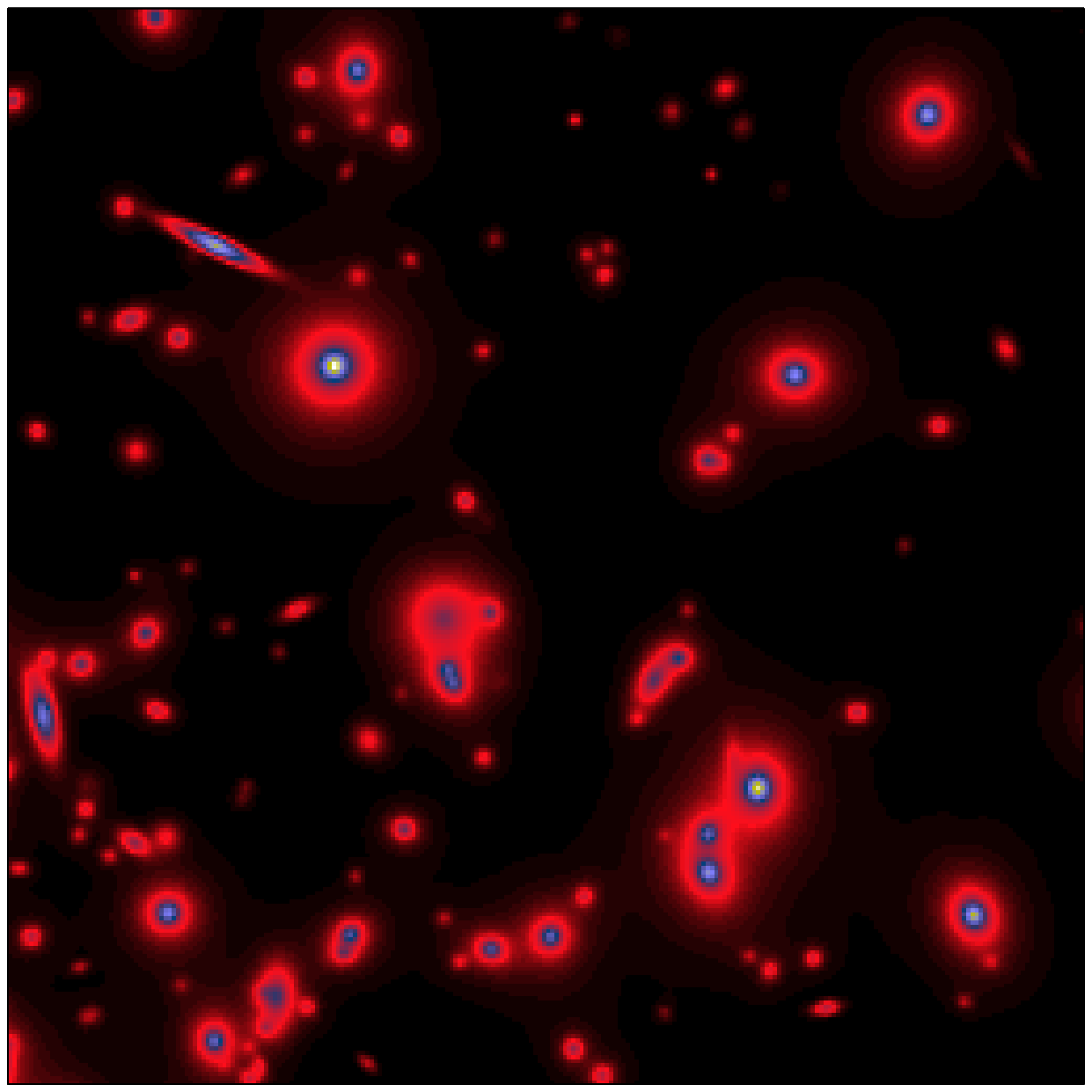} &
      \includegraphics[width=0.3\linewidth]{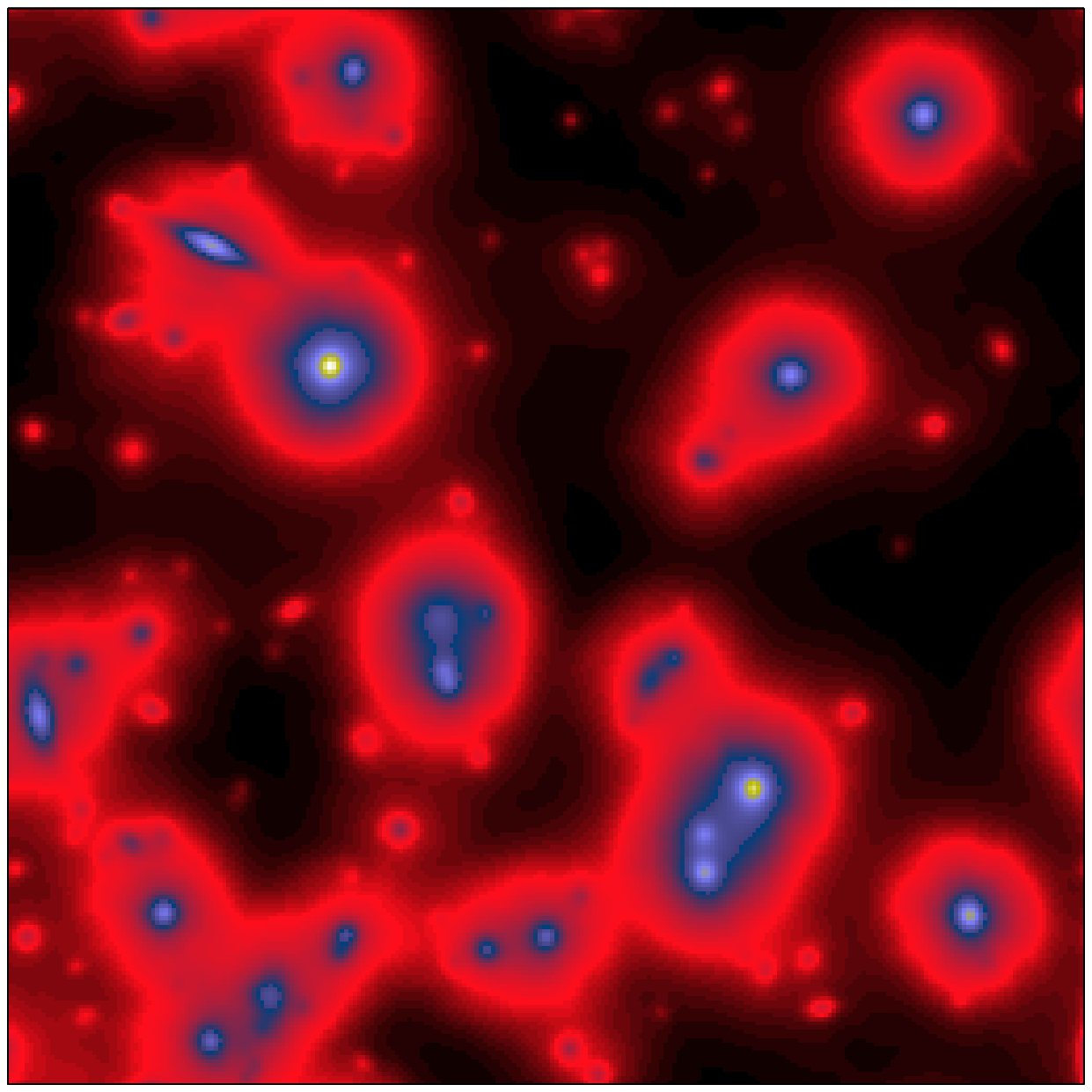} &
      \includegraphics[width=0.3\linewidth]{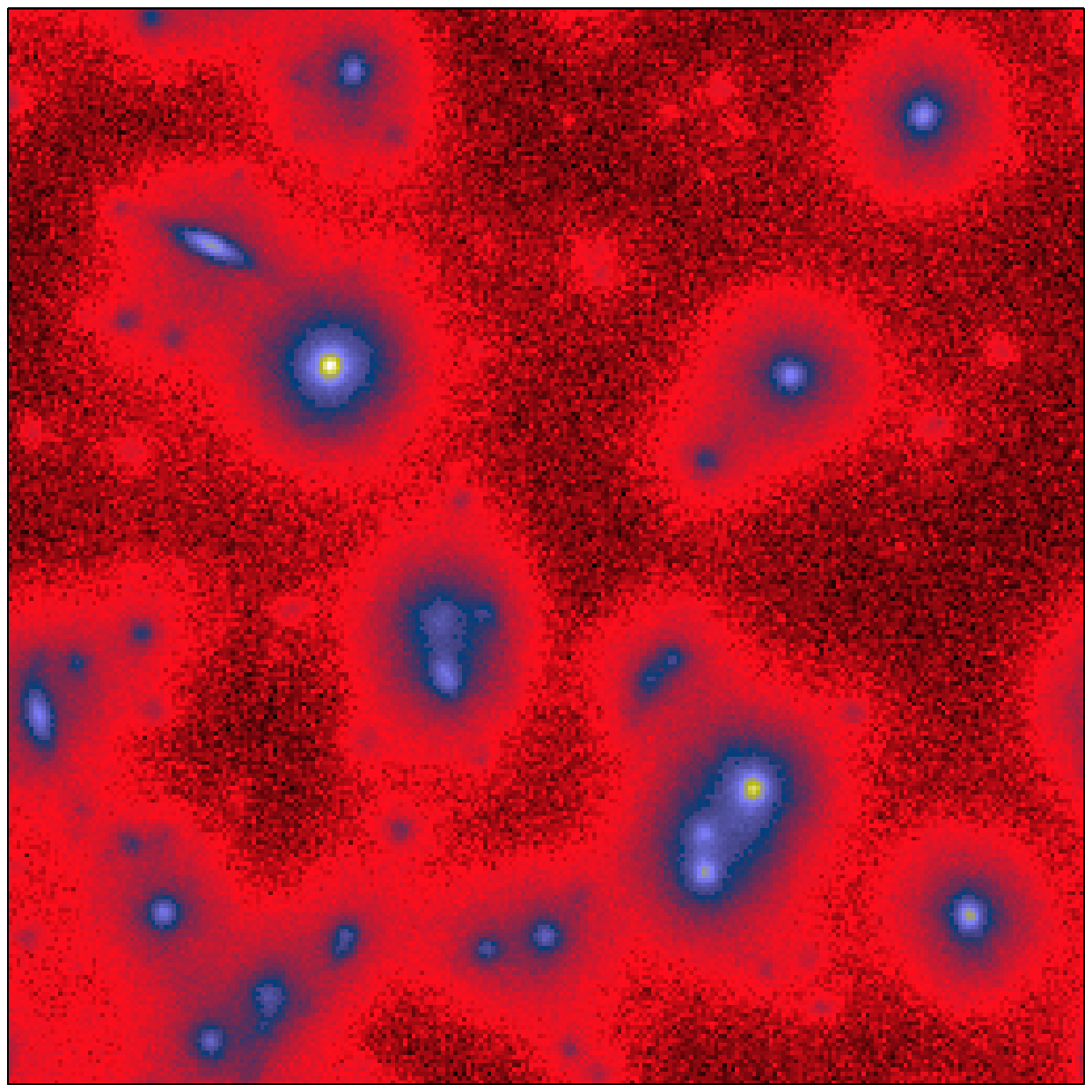} \\
      (a) & (b) & (c)	\\
      \includegraphics[width=0.3\linewidth]{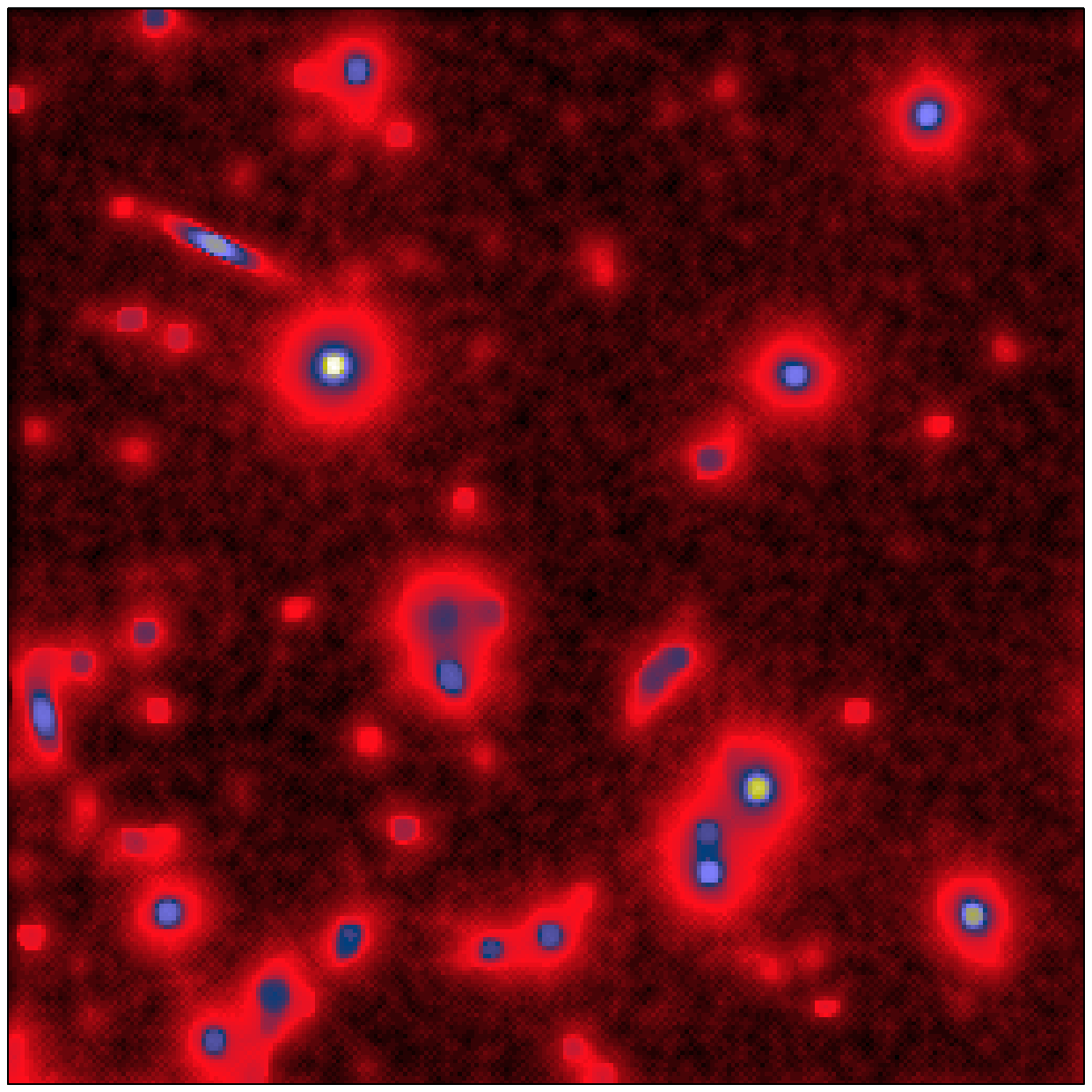} &
      \includegraphics[width=0.3\linewidth]{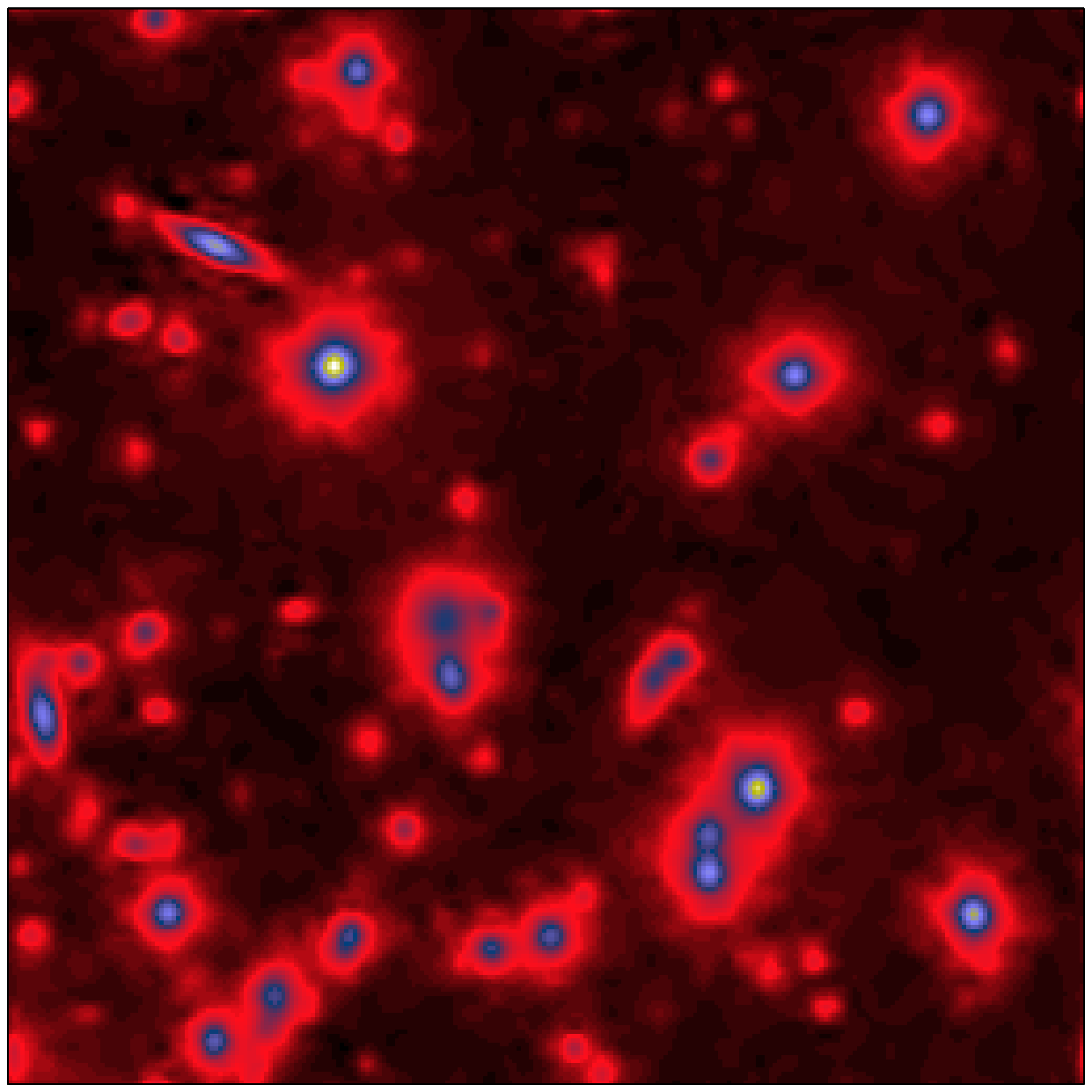} &
      \includegraphics[width=0.3\linewidth]{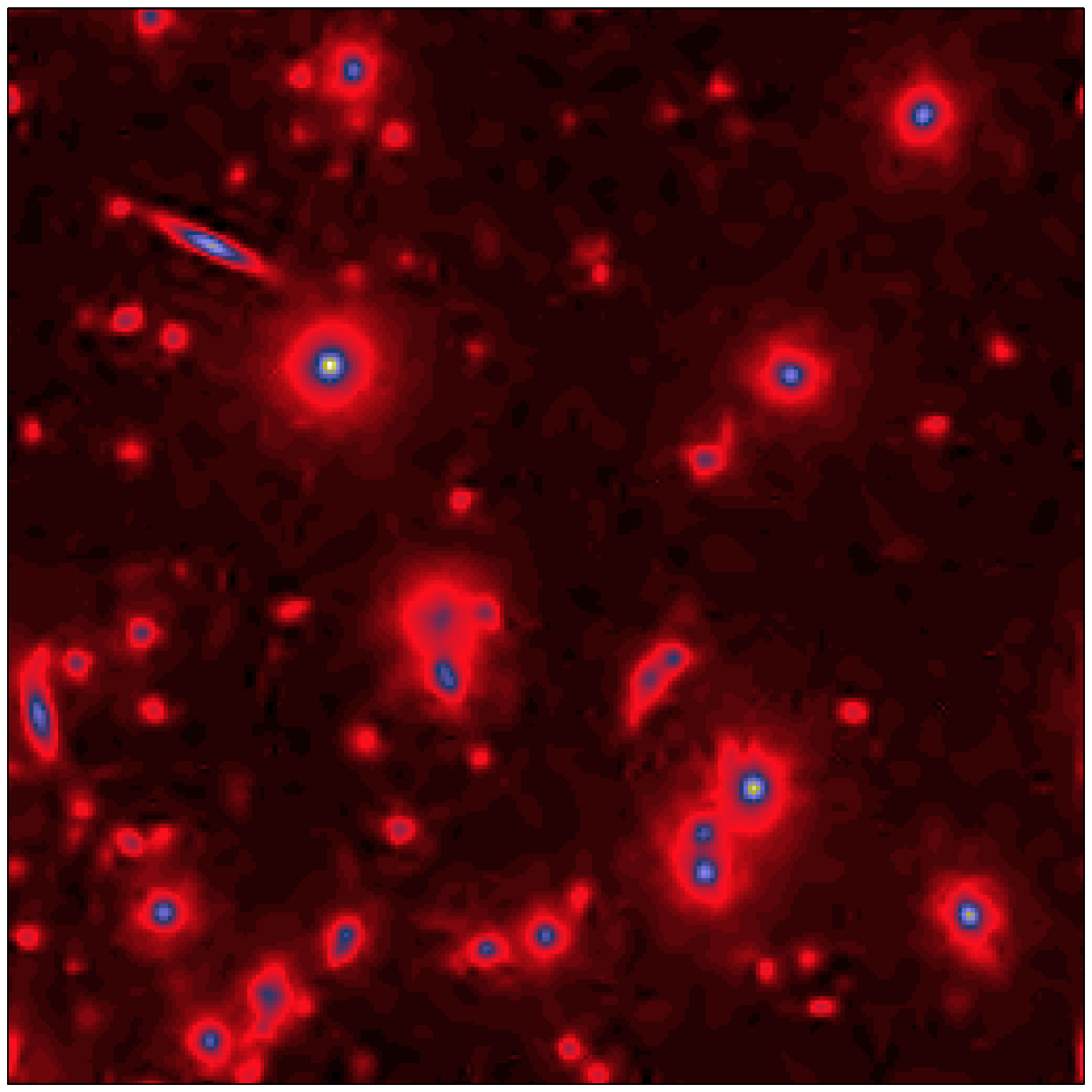} \\
      (d) & (e) & (f) \\
      \includegraphics[width=0.3\linewidth]{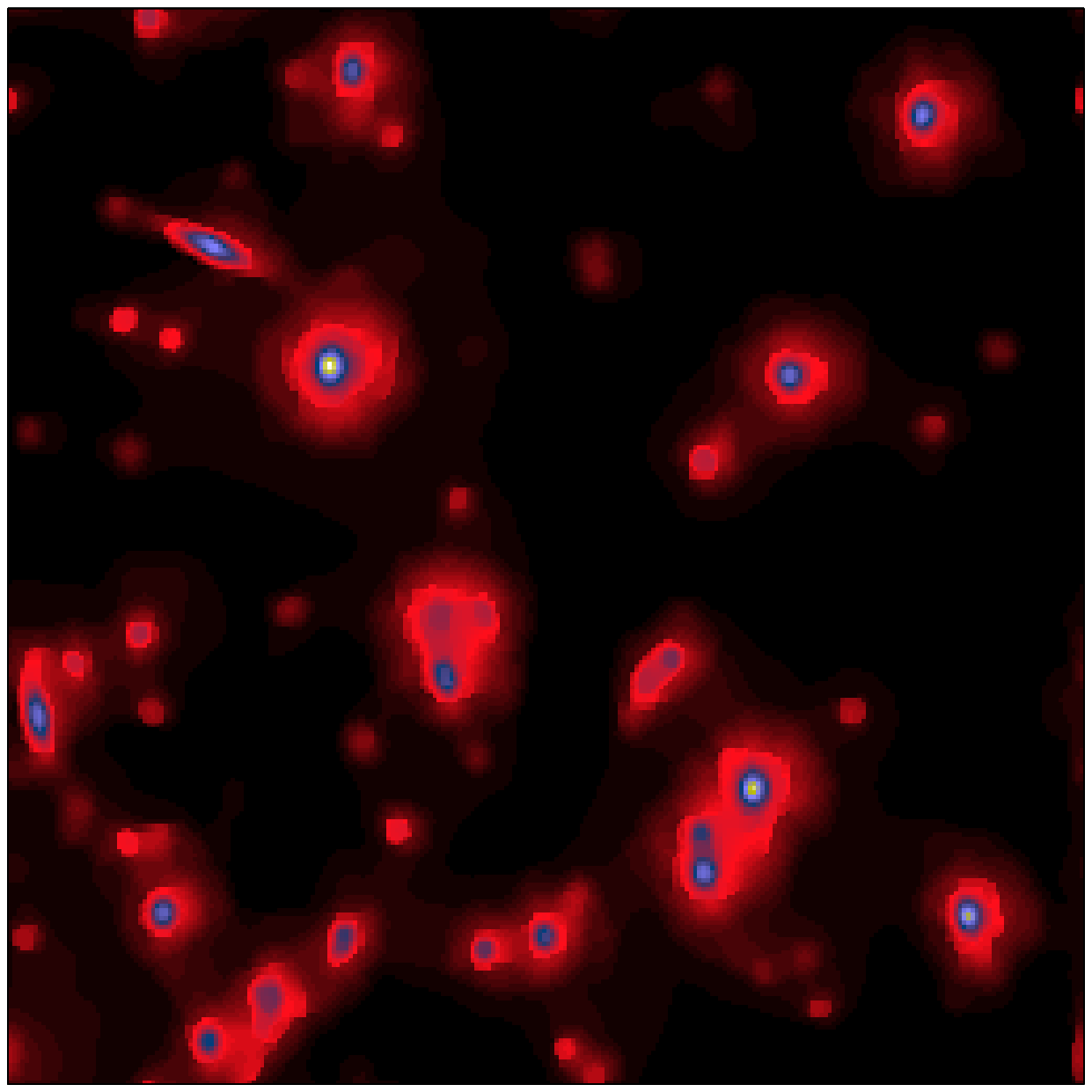} &
      \includegraphics[width=0.3\linewidth]{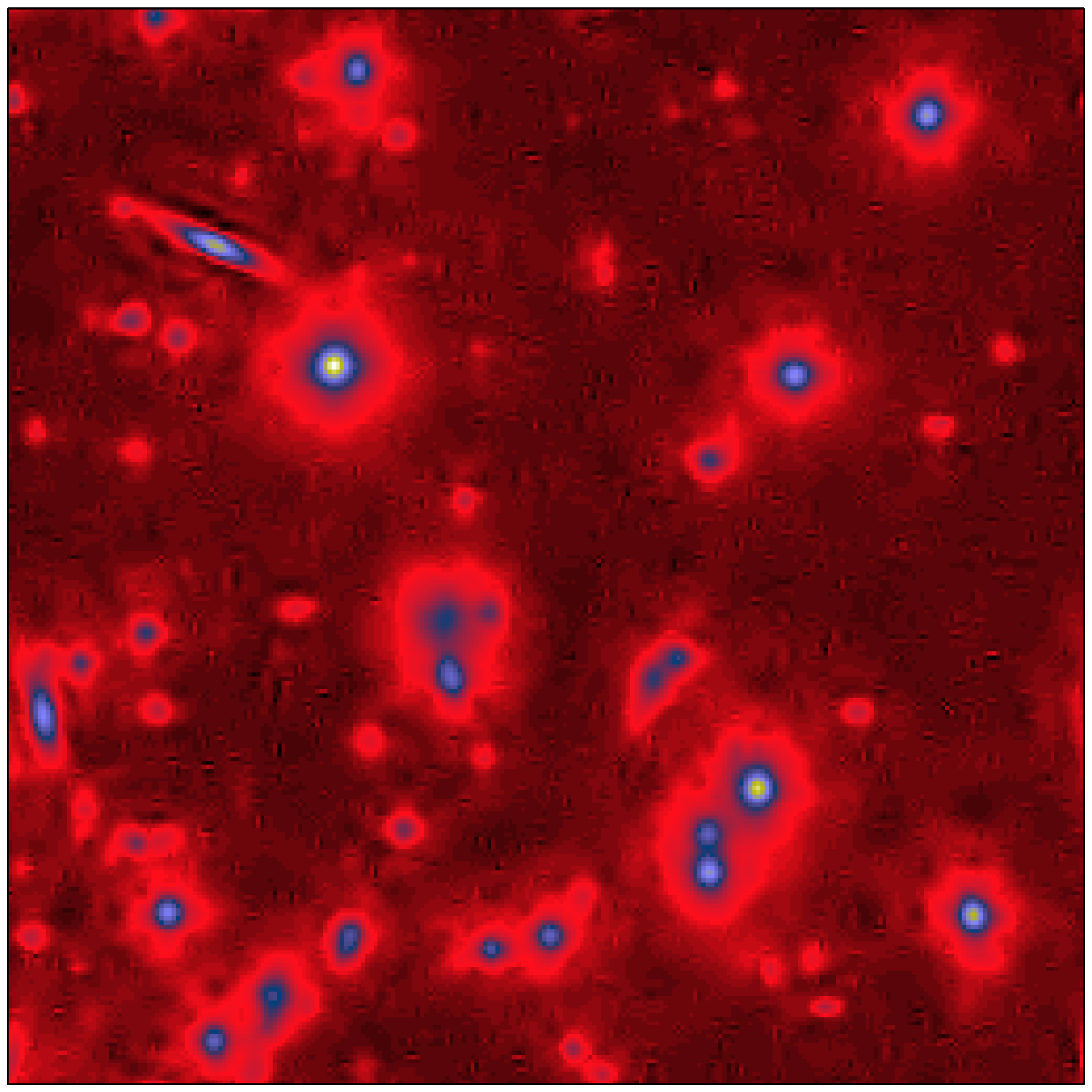}& \\
      (g) & (h) &     \\
    \end{tabular}
  \caption{{Deconvolution of the simulated sky. (a) Original, (b) Blurred,
      (c) Blurred\&noisy, (d) RL-TV \cite{Dey2004}, (e) NaiveGauss \cite{Vonesch2007}, (f)
      RL-MRS \cite{Starck2006}, (g) FTITPR \cite{Willett2004}, (h) Our Algorithm.}}
  \label{fig:sky}
\end{figure}

The same experimental protocol was applied to a simulated Hubble Space Telescope wide
field camera image of a distant cluster of galaxies portrayed in Fig.\ref{fig:sky}(a). We
used the Hubble Space Telescope PSF as given in \cite{Starck2006}. 
The maximum intensity on the blurred image was 5000. For NaiveGauss and our
approach, the dictionary contained the TI-DWT frame. For this image, the RL-MRS is clearly
the best as it was exactly designed to handle Poisson noise for such images. Most faint
structures are recovered by RL-MRS and large bright objects are well deconvolved. Our
approach also yields a good deconvolution result and preserves most faint objects that are
hardly visible on the degraded image. But the background is less clean than the one of
RL-MRS. A this high intensity regime, NaiveGauss provides satisfactory results comparable to ours on the galaxies. 
FTITPR manages to properly recover most significant structures with a very
clean background, but many faint objects are lost. RL-TV gives a deconvolution result
comparable to ours on the brightest objects, but the background is dominated by spurious
faint structures.

\begin{figure}[ht]
  \centering
    \begin{tabular}{@{ }c@{ }c@{ }c@{ }}
      \includegraphics[width=0.3\linewidth]{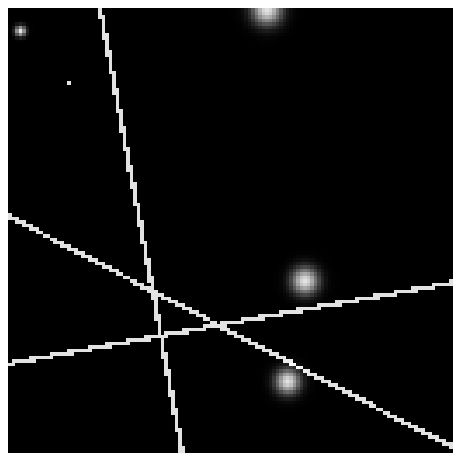} &
      \includegraphics[width=0.3\linewidth]{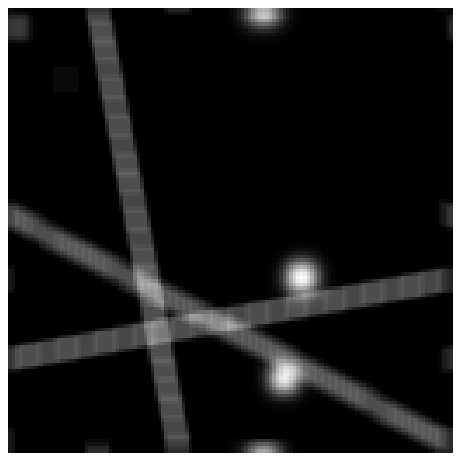} &
      \includegraphics[width=0.3\linewidth]{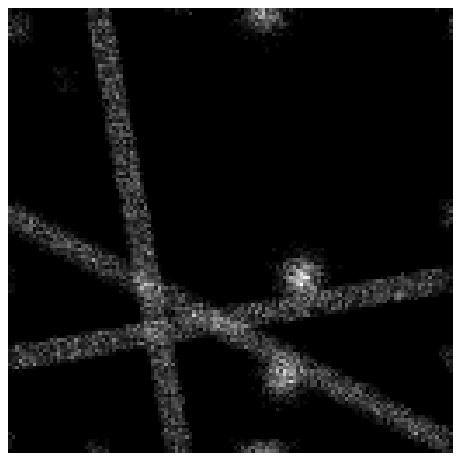} \\
      (a) & (b) & (c)	\\
      \includegraphics[width=0.3\linewidth]{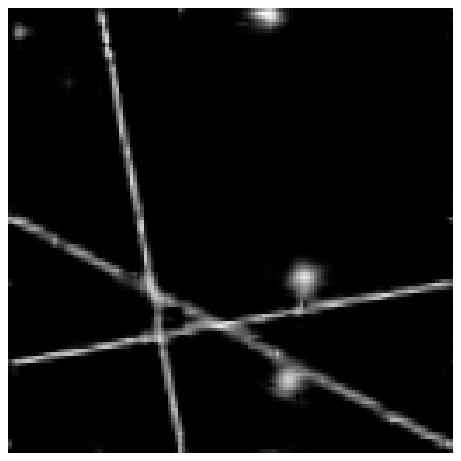} &
      \includegraphics[width=0.3\linewidth]{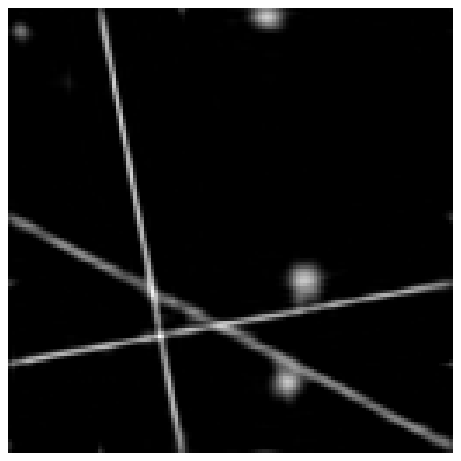} &
      \includegraphics[width=0.3\linewidth]{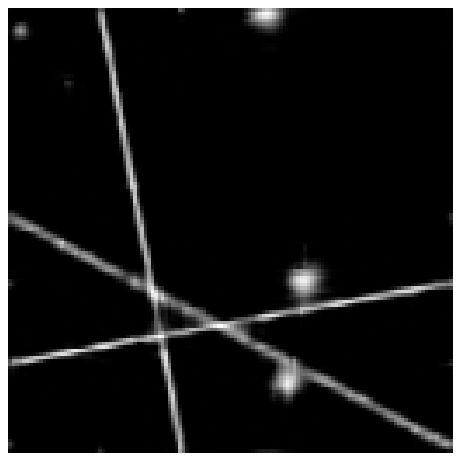} \\
      (d) & (e) & (f)
    \end{tabular}
  \caption{{Impact of the dictionary on deconvolution of the simulated LinesGaussians image with maximum intensity 30. (a) Original, (b) Blurred, (c) Blurred\&noisy, (d) restored with TI-DWT, (e) restored with curvelets, (f) restored with a dictionary containing both transforms.}}
  \label{fig:linesgaussians}
\end{figure}

We also quantified the influence of the dictionary on the deconvolution performance on three test images. We first show in Fig.~\ref{fig:linesgaussians} the results of an experiment on a simulated $128 \times 128$ image, containing point-like sources (upper left), gaussians and lines. In this experiment, the maximum intensity of the original image is 30 and we used the $7 \times 7$ moving-average PSF. The TI-DWT depicted in Fig.~\ref{fig:linesgaussians}(d) does a good job at recovering isotropic structures (point-like and gaussians), but the lines are not well restored. This drawback is overcome when using the curvelet transform as seen from Fig.~\ref{fig:linesgaussians}(e), but as expected, the faint point-like source in the upper-left part is sacrificed. Visually, using a dictionary with both transforms seems to take the best of both worlds, see Fig.~\ref{fig:linesgaussians}(g). 

\begin{figure}[ht]
  \centering
  \hspace*{-1cm}
    \begin{tabular}{@{ }c@{ }c@{ }}
      \includegraphics[width=0.55\linewidth]{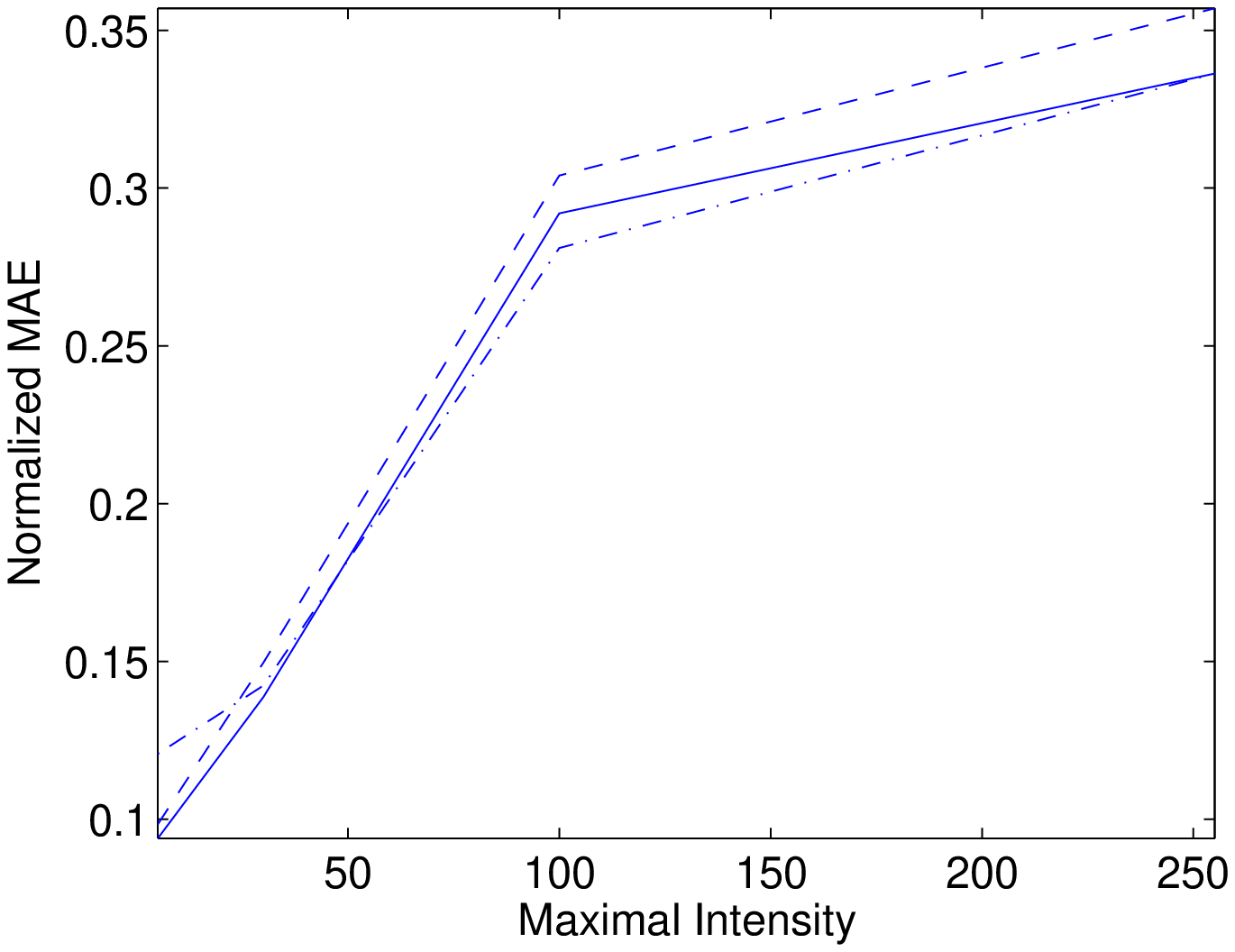} &
      \includegraphics[width=0.55\linewidth]{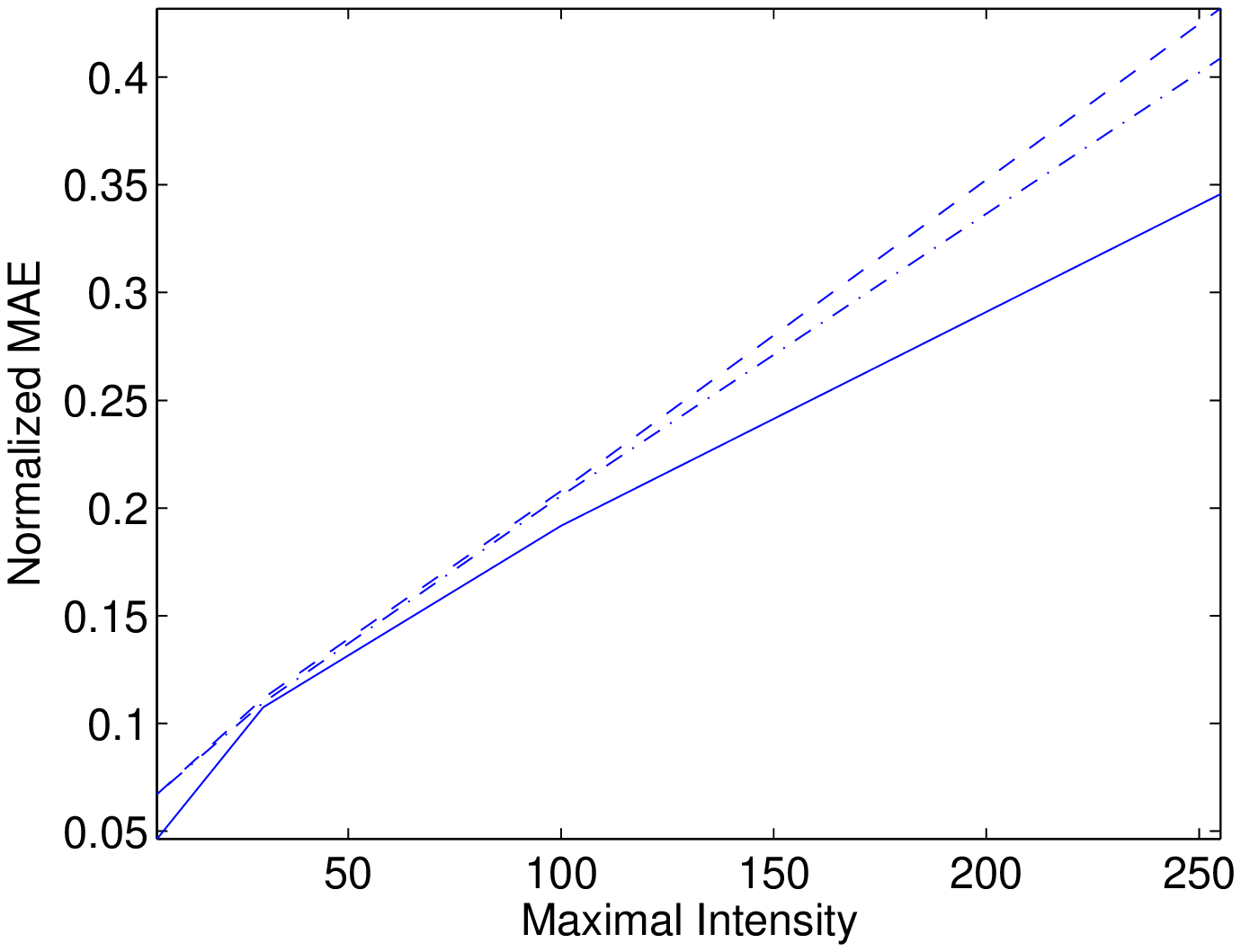} \\
      (a) & (b)
    \end{tabular}
     \hspace*{-1cm} 
     \includegraphics[width=0.55\linewidth]{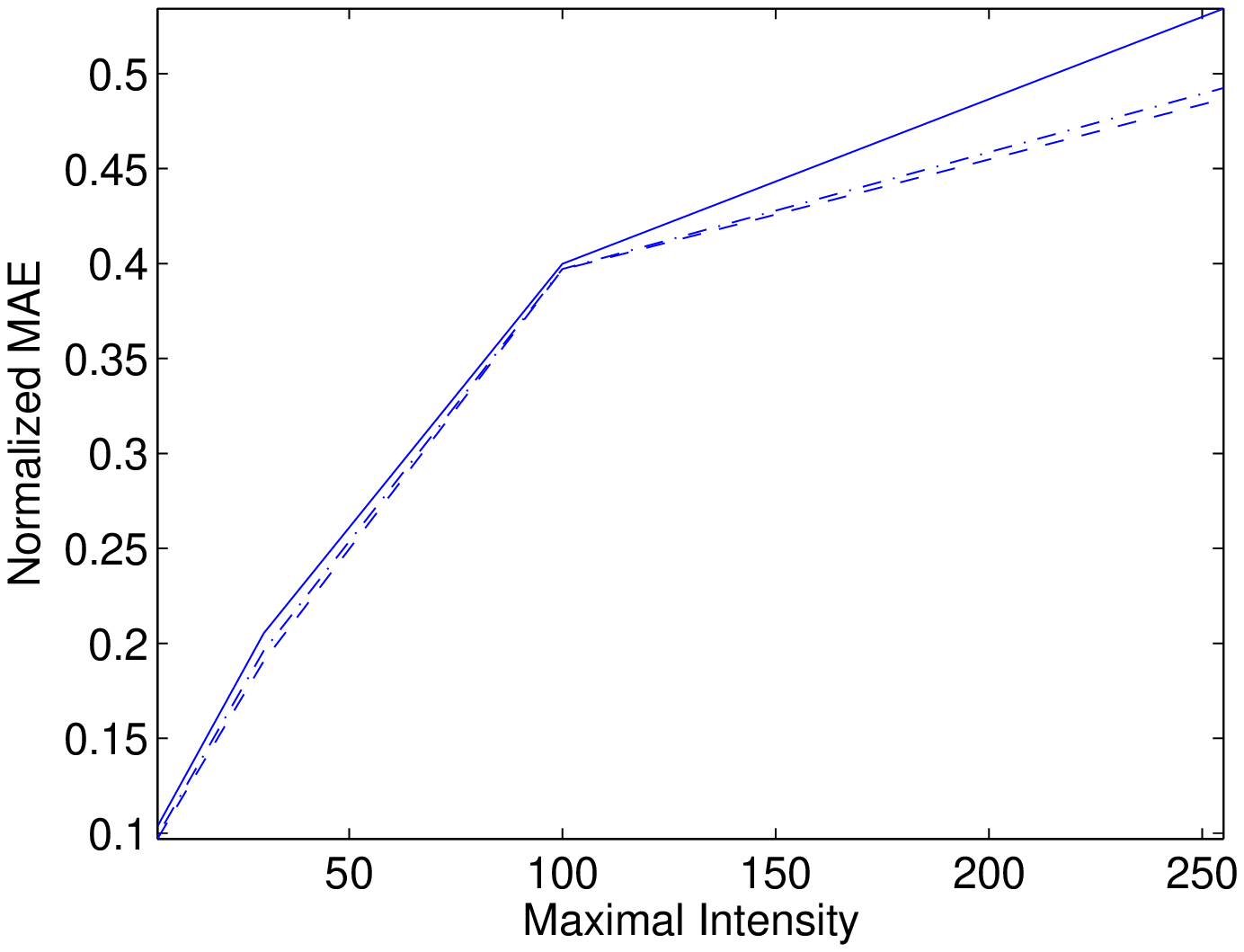} \\
      \hspace*{-1cm} (c)
  \caption{Impact of the dictionary on deconvolution performance as a function of the maximal intensity level for several test images: (a) Neuron phantom, (b) Cell and (c) LinesGaussians images. Solid line indicates the TI-DWT, dashed line corresponds to the curvelet transform, and dashed-dotted line to the dictionary built by merging both wavelets and curvelets.}
  \label{fig:trans}
\end{figure}

Fig.~\ref{fig:trans} shows the MAE{\textemdash}here normalized to the maximum intensity of the original image for the sake of legibility{\textemdash}as a function of the maximal intensity level for three test images: Neuron phantom, Cell and LinesGaussians. As above, three dictionaries were used: TI-DWT (solid line), curvelets (dashed line) and a dictionary built by merging both transforms (dashed-dotted line). For the Neuron phantom, which is piecewise-smooth, the best performance is given by the TI-DWT+curvelets dictionary at medium and high intensities. Even though the differences between dictionaries are less salient at low intensity levels. For the Cell image, which contains many localized structures, the TI-DWT seems to provide the best behavior, especially as the intensity increases. Finally, the behavior observed for the LinesGaussians image is just the opposite to that of the Cell. More precisely, the curvelets and TI-DWT+curvelets dictionaries show the best performance with an advantage to the latter. However, this limited set of experiments does not allow to conclude that a dictionary built by amalgamating several transforms is the best strategy in general. Such a choice strongly depends on the image morphological content.

\subsection{Real data}
\label{sec:real-data}

\begin{figure}[ht]
  \centering
  \footnotesize{
    \begin{tabular}{@{ }c@{ }c@{}}
      \includegraphics[width=0.45\linewidth]{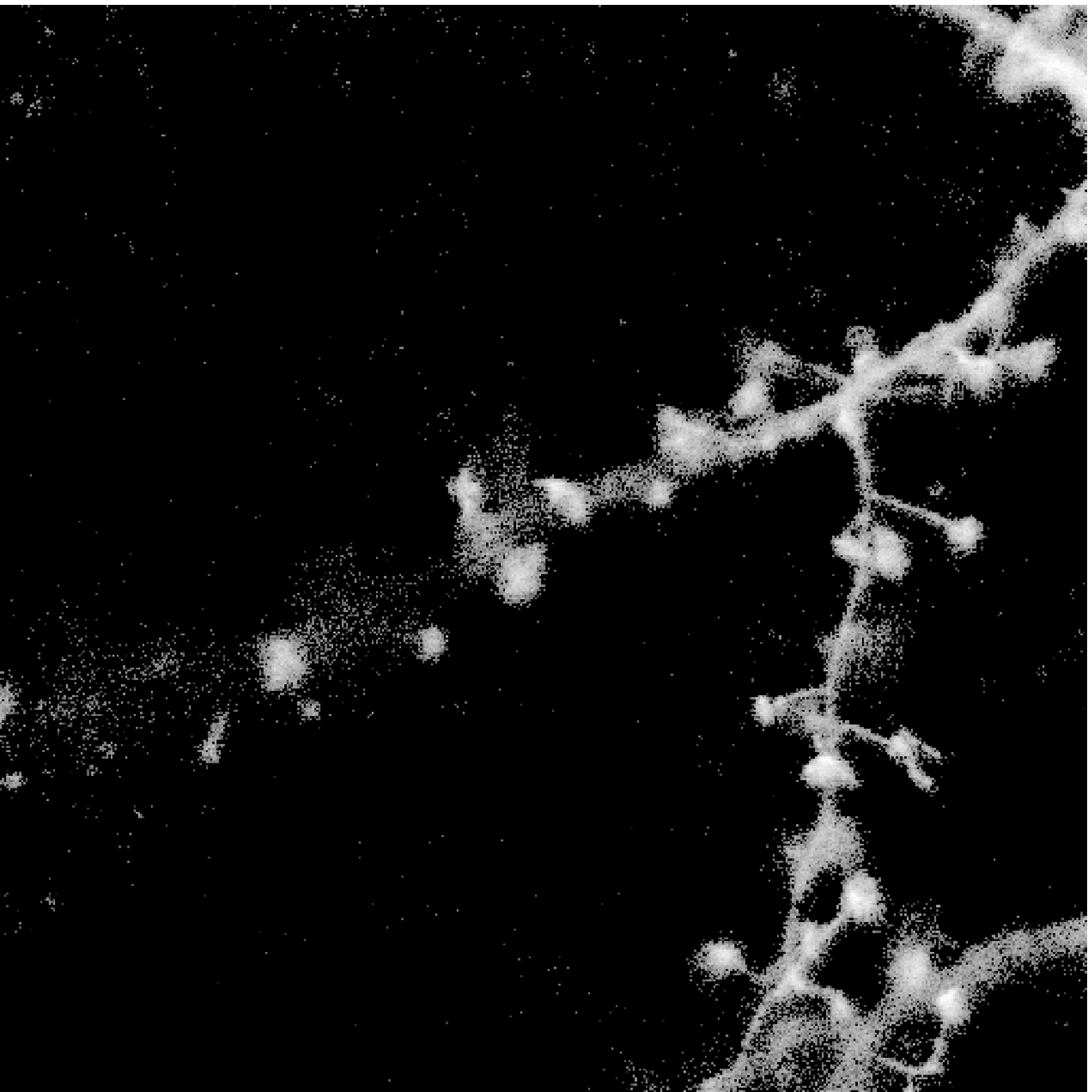} &
      \includegraphics[width=0.45\linewidth]{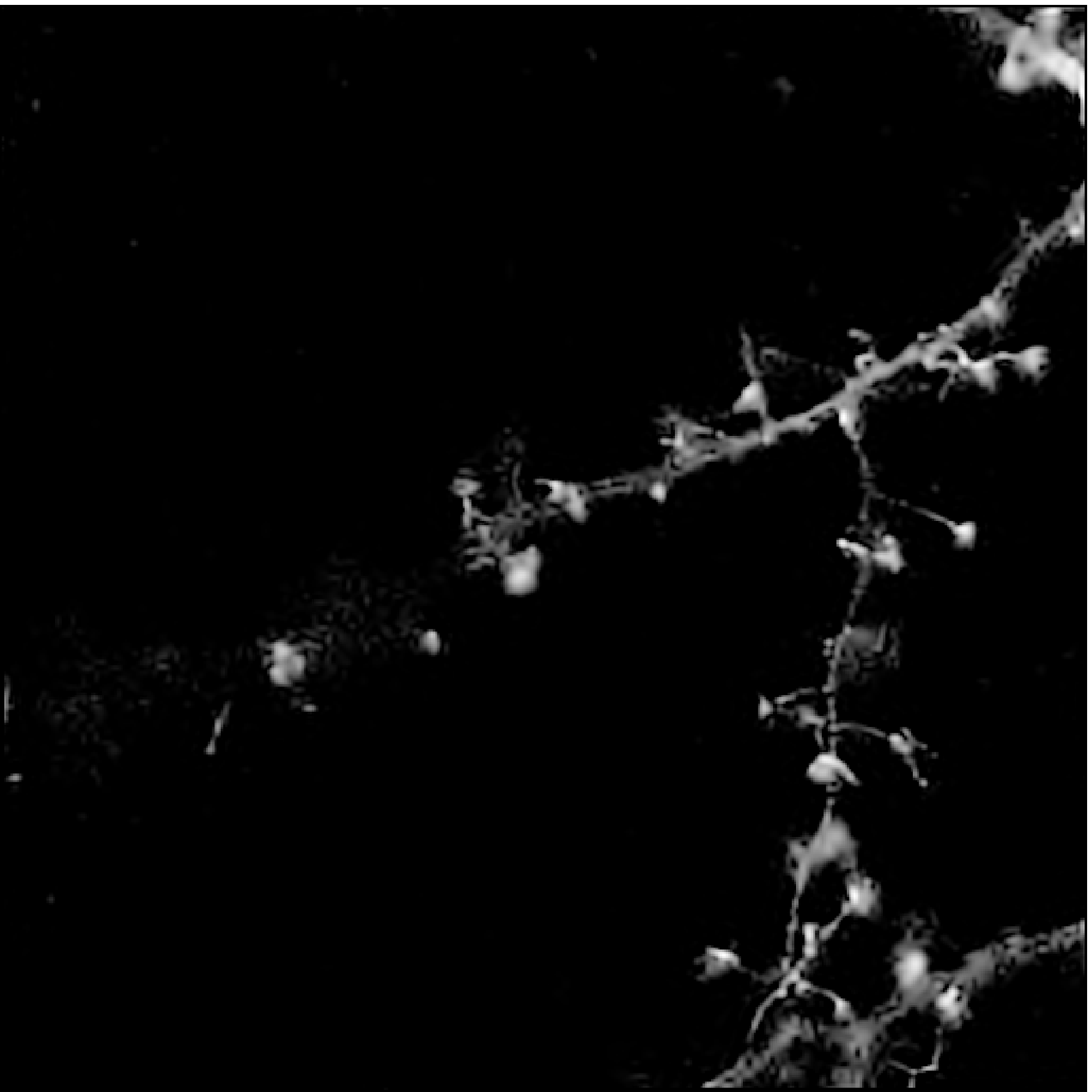} \\
      (a) & (b) \\
    \end{tabular}}
  \caption{\footnotesize{Deconvolution of a real neuron. (a) Original noisy, (b) Restored with our algorithm}}
  \label{fig:real-neuron}
\end{figure}

Finally, we applied our algorithm on a real $512 \times 512$ confocal microscopy image of
neurons.  Fig.~\ref{fig:real-neuron}(a) depicts the observed image\footnote{Courtesy of
  the GIP Cyc\'eron, Caen France.} using the GFP fluorescent protein. The optical PSF of
the fluorescence microscope was modeled using the gaussian approximation described in
\cite{Zhang2006}. Fig.~\ref{fig:real-neuron}(b) shows the restored image using our
algorithm with the wavelet transform. The images are shown in log-scale for better visual
rendering. We can notice that the background has been cleaned and some structures have
reappeared. The spines are well restored and part of the dendritic tree is
reconstructed. However, some information can be lost (see tiny holes). We suspect that
this result may be improved using a more accurate PSF model.

\subsection{Reproducible research}
Following the philosophy of reproducible research \cite{BuckheitDonoho95}, a toolbox is made available freely for download at the first author's webpage \texttt{http://www.greyc.ensicaen.fr/$\sim$fdupe}. This toolbox is a collection of Matlab functions, scripts and datasets for image deconvolution under Poisson noise. It requires at least WaveLab 8.02 \cite{BuckheitDonoho95}. The toolbox implements the proposed algorithms and contains all scripts to reproduce most of the figures included in this paper.

\section{Conclusion}
\label{sec:conclusion}

In this paper, a novel sparsity-based fast iterative thresholding deconvolution algorithm
that takes account of the presence of Poisson noise was presented. The Poisson noise was
handled properly. A careful theoretical study of the optimization problem and
characterization of the iterative algorithm were provided. The choice of the
regularization parameter was also attacked using a GCV-based procedure. Several
experimental tests have shown the capabilities of our approach, which compares favorably
with some state-of-the-art algorithms. Encouraging preliminary results were also obtained
on real confocal microscopy images.

The present work may be extended along several lines. For example, it is worth noting that
our approach generalizes straightforwardly to any non-linearity in \eqref{eq:3} other than
the square-root, provided that the corresponding data fidelity term as in \eqref{eq:7} is convex and has
a Lipschitz-continuous gradient. This is for instance the case if a generalization of the Anscombe VST \cite{Murtagh95} 
is applied to a Poisson plus Gaussian noise, which is a realistic noise model for data obtained from a CCD detector. 
For such a noise, one can easily show similar results to those proved in our work. In this paper, the simple expression 
of the degrees of freedom $df$ was conjectured without a rigorous proof. Deriving the exact analytical expression of $df$, 
if possible, needs further investigation and a very careful analysis that we leave for a future work.
On the applicative side, the extension to 3D to handle
confocal microscopy volumes is under investigation. Extension to multi-valued images is
also an important aspect that will be the focus of future research.

\bibliographystyle{IEEEtran}
\bibliography{spl_bib}

\appendix

\subsection{Proof of Proposition \ref{prop:1}}

\begin{IEEEproof}
  \begin{enumerate}[(i)]
  \item $\dft$ is obviously convex, as $\Phi$ and $\Hd$ are bounded linear operators and
    $f$ is convex.
  \item The computation of the gradient of $\dft$ is straightforward.
  \item For any $\va, \va' \in \Hm$, we have,
    \begin{equation}
      \norm{\nabla\dft(\va) - \nabla\dft(\va')} \leq \norm{\Phi}_2\norm{\Hd}_2
      \norm{\nabla F\circ\Hd\circ\Phi(\va) - \nabla F\circ\Hd\circ\Phi(\va')}.
    \end{equation}
    The function $-\frac{z_i}{\sqrt{\eta_i + 3/8}} + 2$ is one-to-one increasing on
    $[0,+\infty)$ with derivative uniformly bounded above by
    $\frac{z_i}{2}\parenth{8/3}^{3/2}$. Thus,
    \begin{eqnarray}
      \norm{\nabla F\circ\Hd\circ\Phi(\va) - \nabla F\circ\Hd\circ\Phi(\va')}
      &\leq& \parenth{\frac{8}{3}}^{\tfrac{3}{2}} \frac{\norm{z}_{\infty}}{2}
      \norm{\Hd\circ\Phi(\va) - \Hd\circ\Phi(\va')} \nonumber\\
      &\leq& \parenth{\frac{8}{3}}^{\tfrac{3}{2}} \frac{\norm{z}_{\infty}}{2} \norm{\Phi}_2\norm{\Hd}_2 \norm{\va - \va'} ~ .
    \end{eqnarray}
    Using the fact that $\norm{\Phi}^2_2 = \norm{\Phi\Phi\Tr}_2=c$ for a tight frame, and
    $z$ is bounded since $y \in \ell_\infty$ by assumption, we conclude that $\nabla\dft$
    is Lipschitz-continuous with the constant given in \eqref{eq:22}.
  \end{enumerate}
\end{IEEEproof}

\subsection{Proof of Proposition \ref{prop:2}}

\begin{IEEEproof}
  The existence is obvious because $J$ is coercive. If $\Phi$ is an orthobasis and
  $\mathrm{ker}\parenth{\Hd} = \emptyset$ then $\dft$ is strictly convex and so is $J$
  leading to a strict minimum. Similarly, if $\psi$ is strictly convex, so is $\reg$, hence
  $J$.
\end{IEEEproof}

\subsection{Proof of Lemma \ref{lemma:2}}

\begin{IEEEproof}  
  \begin{enumerate}
  \item Let $g: \gamma \mapsto \tfrac{1}{2}\norm{\va - \gamma}^2 + \lambda
    \Psi(\gamma)$. From Definition~\ref{def:1}, $\prox_{\lambda\Psi}(\va)$ is the unique
    minimizer of $g$, whereas $\prox_{f_2}(\va)$ is the unique minimizer of $g +
    \imath_{\C'}$. If $\va \in \C'$, then $\prox_{f_2}(\va)$ is also the unique minimizer
    of $g$ as obviously $\imath_{\C'}(\va) = 0$ in this case. That is, $\prox_{f_2}(\va) =
    \prox_{\lambda\Psi}(\va)$.
 

  \noindent
  
  \item Let's now turn to the general case. We have to find the unique solution to the following minimization problem:
  \begin{equation*}
    \prox_{\reg}(\va) = \argmin_{\gamma} ~ g(\gamma) + \imath_{\C}\circ\Phi(\gamma) = \argmin_{\gamma \in \C'} ~ g(\gamma).
  \end{equation*}
  As both $\imath_{\C}$ and $g \in \Gamma_0\parenth{\R^L}$ but non-differentiable, we use
  the Douglas-Rachford splitting method \cite{Eckstein92,Combettes04}. This iteration is
  given by:
  \begin{equation}
    \gamma^{t+1} = \gamma^t + \nu_t\parenth{\rprox_{\lambda\Psi + \tfrac{1}{2}\norm{. - \va}^2}\circ\rprox_{\imath_{\C'}} - \I}(\gamma^t).
  \end{equation}
  where the sequence $\nu_t$ satisfies the condition of the lemma. From \cite[Corollary
  5.2]{Combettes04}, and by strict convexity, we deduce that the sequence of iterates
  $\gamma^t$ converges to a unique point $\gamma$, and $\Prj_{\C'}(\gamma)$ is the
  unique proximity point $\prox_{f_2}(\va)$.
  
  It remains now to explicitly express $\prox_{\lambda\Psi + \tfrac{1}{2}\norm{. -
      \va}^2}$ and $\prox_{\imath_{\C'}}$. $\prox_{\lambda\Psi + \tfrac{1}{2}\norm{. -
      \va}^2}$ is the proximity operator of a quadratic perturbation of $\lambda\Psi$,
  which is related to $\prox_{\lambda\Psi}$ by:
  \begin{equation}
    \prox_{\lambda\Psi+\tfrac{1}{2}\norm{. - \va}^2} ( . ) = \prox_{\tfrac{\lambda}{2}\Psi} \left(
    \frac{\va + .}{2}\right).
  \end{equation}
  See \cite[Lemma 2.6]{Combettes2005}.
  
  Using \cite[Proposition 11]{Combettes2007a}, we have
  \begin{eqnarray}
    \prox_{\imath_{C} \circ \Phi} &=& \I + c^{-1}\Phi\Tr \circ \parenth{\Prj_\C - \I} \circ \Phi \nonumber \\
    				  &=& c^{-1}\Phi\Tr \circ \Prj_\C \circ \Phi + (\I - c^{-1}\Phi\Tr\Phi).
  \end{eqnarray}
  \end{enumerate}
  This completes the proof.
\end{IEEEproof}

\subsection{Proof of Theorem~\ref{th:2}}

\begin{IEEEproof}
  The most general result on the convergence of the forward-backward algorithm is is due
  to \cite[Theorem 3.4]{Combettes2005}.  Hence, combining this theorem with Lemma
  \ref{lemma:2}, Lemma \ref{th:3} and Proposition \ref{prop:1}, the result follows.
\end{IEEEproof}

\end{document}